\documentclass[sn-mathphys-num]{sn-jnl}

\usepackage{graphicx}%
\usepackage{multirow}%
\usepackage{amsmath,amssymb,amsfonts}%
\usepackage{amsthm}%
\usepackage{mathrsfs}%
\usepackage[title]{appendix}%
\usepackage{xcolor}%

\usepackage{textcomp}%
\usepackage{manyfoot}%
\usepackage{booktabs}%
\usepackage{algorithm}%
\usepackage{algorithmicx}%
\usepackage{algpseudocode}%
\usepackage{listings}%
\usepackage{tikz}
\usetikzlibrary{decorations.pathmorphing}
\usepackage[utf8]{inputenc}
\usepackage{amsmath}
\usepackage{natbib}
\usepackage{color}
\usepackage{mathrsfs}
\usepackage{bigints}

\usepackage{amssymb}
\usepackage{comment}
\usepackage{float}
\usepackage{caption}
\usepackage[usenames,dvipsnames]{pstricks}
\usepackage{epsfig}
\usepackage{pst-grad} 
\usepackage{pst-plot} 
\usepackage[space]{grffile} 
\usepackage{etoolbox} 
\makeatletter 
\patchcmd\Gread@eps{\@inputcheck#1 }{\@inputcheck"#1"\relax}{}{}
\makeatother

\begin{document}

\title[Article Title]{Classical general relativity effects by magnetars with massive quadrupole, angular momentum and a magnetic dipole}


\author*[1,2]{\fnm{Alexander} \sur{Mora-Chaverri}}\email{alexander.morachaverri@ucr.ac.cr}

\author[2,3]{\fnm{Edwin} \sur{Santiago-Leandro}}\email{edwin.santiago@ucr.ac.cr}
\equalcont{These authors contributed equally to this work.}

\author[1,2]{\fnm{Francisco} \sur{Frutos-Alfaro}}\email{francisco.frutos@ucr.ac.cr}
\equalcont{These authors contributed equally to this work.}

\affil*[1]{\orgdiv{School of Physics}, \orgname{University of Costa Rica}, \orgaddress{\street{San Pedro}, \city{Montes de Oca}, \postcode{11801}, \state{San Jos\'e}, \country{Costa Rica}}}

\affil[2]{\orgdiv{School of Physics}, \orgname{University of Costa Rica}, \orgaddress{\street{San Pedro}, \city{Montes de Oca}, \postcode{11801}, \state{San Jos\'e}, \country{Costa Rica}}}

\affil[3]{\orgdiv{School of Physics}, \orgname{University of Costa Rica}, \orgaddress{\street{San Pedro}, \city{Montes de Oca}, \postcode{11801}, \state{San Jos\'e}, \country{Costa Rica}}}


\abstract{In this contribution, we obtain classical tests of general relativity using the Hartle-Thorne metric endowed with magnetic dipole and electric charge. This metric represents the approximate stationary spacetime of a massive object with the other characteristics mentioned. These tests are light deflection, time delay, periastron precession, and gravitational redshift. We also provide numerical estimates for real magnetars and magnetar candidates from the McGill magnetar catalog, the millisecond pulsar PSR B1257+12 and for the Sun in low-activity cycles. Our results find that, although the magnetic dipole moment contribution tends to be negligible compared to the total amount, its comparison to the massive quadrupole moment and rotational contributions varies from one classical test to the next. For light deflection, the magnetic dipole contribution is about 2 orders of magnitude smaller, compared to the rotational contribution. The magnetic dipole moment contribution is present, but is about 6 orders of magnitude smaller than the second-order rotational contribution to the periastron precession, 5 orders of magnitude smaller for the time delay, and negligible within the approximation presented for the gravitational redshift. The magnetic dipole contribution for the calculations made with PSR B1257+12 was also negligible, but the rotational and quadrupole moment contributions were more significant, which makes the argument for possible future detection stronger than the magnetar case. The rotation, massive quadrupole moment and magnetic dipole contributions for the Sun turned out to be negligible as well.}




\maketitle

\section{Introduction}\label{sec1}

Since the formulation of General Relativity at the beginning of the 20th century, there has been a significant motivation to find diverse solutions to Einstein's equations by incorporating more complex physical phenomena. This effort was further driven by Kerr's discovery in 1963 of a solution describing a rotating massive object \cite{Kerr}. A few years later, Bonnor derived an exact solution representing a static massive source with a magnetic dipole moment. However, this solution does not reduce to the well-known Schwarzschild metric when the magnetic field is zero \cite{Bonnor}. Subsequently, Martin and Pritchett found an approximate magnetic dipole solution to the Einstein-Maxwell equations (EME) \cite{MP}.

\noindent
In 1968, Hartle and Thorne (HT) developed an approximate metric incorporating mass, rotation, and a massive quadrupole moment, which models neutron stars (NS), white dwarfs, and massive stars with slow rotation and small deformation \cite{Hartle1968}. In 1993, Manko constructed a spacetime with both magnetic field and rotation, but this model lacked massive deformation \cite{Manko93}. Deformation of a spherical object can arise due to both rotation and magnetic stress. Konno specifically analyzed the relativistic effects of magnetic stress, applying them to magnetars \cite{Konno99, Konno20}. Yagi, in 2014, studied deformation using a higher-order multipole expansion, excluding magnetic stress \cite{Yagi}.

\noindent
In 2006, Pachón et al. developed an exact metric incorporating rotation, magnetic dipole, electric charge, mass quadrupole moment, and current octupole moment \cite{Pachon}. Frutos-Alfaro and Soffel later obtained a static HT spacetime including the second-order contribution of the massive quadrupole \cite{F-S}. In 2016, an approximate Kerr-like metric was developed by expanding the Erez-Rosen metric up to second order in the mass quadrupole \cite{Frutos16, Frutos1}. Transformations between the series-expanded Erez-Rosen metric and a static HT metric, as well as between the approximate Kerr-like metric and the HT metric, were identified. These transformations suggest that the resulting metrics may have an interior counterpart, as the original HT metric does. The significance of these transformations has been demonstrated by Bini et al. \cite{Bini}.  This metric differs from ours because it does not contain the second-order mass quadrupole term. In 2024, Frutos-Alfaro perturbatively extended this metric to include electric charge and a magnetic dipole moment, which reduces to the Reissner–Nordström metric when both the magnetic field and angular momentum vanish \cite{Frutos2}. This HT-like metric provides an interesting model for describing compact objects. It is worth mentioning that in 2016, Steil wrote a master's thesis in which he developed both the internal and external model of a compact object based on the HT metric in which he included the electric charge and the magnetic dipole \cite{Steil}.
Although spacetimes with magnetic dipole moments are seldom used in astrophysical simulations due to their complexity, the HT-like metric offers a simpler model, enabling analytical expressions for phenomena such as the classical tests of General Relativity, including light deflection, periastron precession, time delay, and gravitational redshift.

\noindent
Previous studies refined these classical tests by considering the contributions of mass \cite{Bodenner}, angular momentum \cite{Subra}, mass quadrupole \cite{Suliyeva, Arce}, and magnetic dipole moments. While light deflection contributions from magnetic dipoles have been analyzed using different formalisms, these studies lack estimates for other classical tests and do not account for object rotation or the coupling between magnetic fields and spacetime. For example, geometric optics approaches (eikonal equation, Gauss-Bonnet theorem) or approximate spacetimes with non-rotating dipole moments have been used to study nonlinear electrodynamic effects on light deflection \cite{Beissen, Denisov, Kim, KL}.

\noindent
In this work, we derive expressions incorporating previously studied quantities, including the electric charge contribution, using the HT-like metric. We compute the magnetic dipole contribution and provide numerical estimates using data from the McGill Magnetar Catalog \cite{OK}, considering the coupling between magnetic fields and spacetime. All calculations were performed using Python scripts and REDUCE, a computer algebra system similar to Mathematica \cite{REDUCE}. These codes are available upon request.

\noindent
Additionally to the magnetar candidates, we analyze and perform calculations for the millisecond pulsar PSR B1257+12, which hosts three exoplanets \cite{Wolszczan}. This enables comparisons between calculations for magnetars and millisecond pulsars, as these objects have distinct parameters, such as periods and magnetic fields (see Sec. \ref{num-est}). Pulsars with exoplanets are an interesting phenomenon to study classical tests because it is more likely in these systems that contributions due to angular momentum, deformation or magnetic field would play a significant role. However, it is also important to mention that pulsar planets are rare \cite{Martin}. We also perform the same calculations for the Sun in low cycles of activity.

\noindent
For initial calculations, we utilized the Gutsunaev-Manko (GM) metric \cite{GM}. However, this metric exhibits a pathology: in the weak-field approximation, its Taylor series contains unphysical terms due to the parameter representing the magnetic dipole. Consequently, we adopted the HT-like metric instead (see Appendix C).

\noindent
This paper is organized as follows. Section 2 describes the HT-like metric. Section 3 discusses the classical tests of General Relativity, with expressions derived in the equatorial plane of the HT-like metric. Section 4 presents the magnetic field for a local non-rotating observer (LNRO) in the HT-like metric \cite{Frutos2}. This section also relates the magnetic dipole moment ($\mu$) to realistic objects, such as magnetars, PSR B1257+12 and the Sun during low-activity cycles, and provides tables with classical test estimates for these celestial objects. Conclusions are discussed in Section 5. Throughout this article, we use the geometrized unit system ($G=c=1$) for algebraic expressions, with specified units for numerical quantities.

\section{Hartle-Thorne-like metric}

The HT-like metric describes the spacetime of an object that has mass, angular momentum, electric charge, magnetic dipole, and mass quadrupole \cite{Frutos2}. It is an approximate solution of the EME. This metric in spherical coordinates is

\begin{equation}
\label{LWP}
ds^2 = -V dt^2 + 2W dt d\varphi + X dr^2 + Y d \theta^2 + Z d \varphi^2 ,
\end{equation}

\noindent 
where the metric potentials are

\begin{align}
V &= \Big(1-2Mu+Q_e^2 u^2 - 2/3J^2u^4\Big)e^{2\psi_1} , \nonumber \\
W &= -2Ju\sin^2{\theta}-JQP^1_3u^4\sin{\theta} + W_\mu , \nonumber \\
X &= \Big(1-2Mu+Q_e^2 u^2 + 2J^2u^4\Big)^{-1}e^{-2\psi_2} , \\
Y &=  u^{-2} e^{-2\psi_3} , \nonumber \\
Z &= Y \sin^2{\theta} , \nonumber 
\end{align}

\noindent
with the following functions

\begin{align}
\psi_1 &= QP_2u^3+3MQP_2u^4 - \frac{2}{3}J^2P_2u^4 + V_\mu , \nonumber \\
\psi_2 &= QP_2u^3+3MQP_2u^4 - 8J^2P_2u^4 + \frac{1}{24}Q^2(16P^2_2+16P_2-77)u^6 + X_\mu , \nonumber  \\
\psi_3 &= QP_2u^3+\frac{5}{2}MQP_2u^4 - \frac{1}{2}J^2P_2u^4 + \frac{1}{72}Q^2(28P^2_2-8P_2+43)u^6 + Y_\mu , \nonumber \\
V_\mu &=  \frac{1}{2}\mu^2u^4\cos^2{\theta} , \\ 
W_\mu&= \mu Q_e u^2 \sin^2{\theta} , \nonumber \\
X_\mu &=-\frac{1}{3}\mu^2(P_2-1)u^4 ,\nonumber\\
Y_\mu&= \frac{1}{6}\mu^2P_2u^4 , \nonumber 
\end{align}

\noindent
where $M, \, J, \, Q, \, \mu$ and $Q_e$ are  total mass, angular momentum, massive quadrupole moment, magnetic dipole moment and electric charge, respectively and $u=1/r$.

\noindent 
The functions $P_2$ and $P^{1}_{3}$ are Legendre polynomials

\begin{align}
P_2 &= \frac{1}{2} (3 \cos^2{\theta} - 1) , \nonumber \\
P^{1}_{3} &= (5P_2 + 1)\sin{\theta} . \nonumber 
\end{align}

\noindent
The electromagnetic four-potential is given by

\begin{align}
\label{Aphi}
A_t &= -Q_eu-J\mu u^4 \cos^4{\theta}-QQ_eP_2u^4,  \\
A_\varphi &= \Big(\mu u - \frac{3}{2}J Q_e u^2 + \frac{3}{2} M \mu u^2 + \frac{1}{4}\mu Q (P_2+1)u^4\Big) \sin^2{\theta} . \nonumber
\end{align}

\noindent
The electromagnetic field components (EMFC) for a locally non-rotating observer (LNRO) can be determined by

\begin{align}
    E_r&=\frac{1}{\sqrt{X\Big(VZ+W^2\Big)}}\Big[\partial_r A_t - \frac{W}{\sqrt{Z}}\partial_r A_\varphi\Big], \nonumber \\
    E_\theta&=\frac{1}{\sqrt{Y\Big(VZ+W^2\Big)}}\Big[\partial_\theta A_t - \frac{W}{\sqrt{Z}}\partial_\theta A_\varphi\Big] , \\
    H_r&= \frac{1}{\sqrt{YZ}}\partial_\theta A_\varphi , \nonumber \\
    H_\theta&= -\frac{1}{\sqrt{XZ}}\partial_r A_\varphi , \nonumber 
\end{align}

\noindent
Using Eq.(\ref{Aphi}) the EMFC are

\begin{align}
E_r &= Q_eu^2+4(J\mu+QQ_e)P_2 u^5 , \nonumber \\
E_\theta &= \frac{3}{2}(2J\mu+Q Q_e)u^5 \sin({2\theta}) , \\
H_r &= (2\mu u^3+3(\mu M - J Q_e)u^4 + 5\mu Q P_2 u^6)\cos{\theta} , \nonumber \\
H_\theta &= (\mu u^3+(2\mu M - 3J Q_e)u^4 + \mu Q (3P_2+1) u^6)\sin{\theta} . \nonumber
\end{align}

\section{Classical tests of general relativity}

In general relativity, there are many interesting effects that can be observed in the solar system or in other systems. Some of these effects are known as the classical tests of general relativity: periastron precession, light deflection, gravitational redshift, and time delay. These phenomena cannot be explained using Newton’s law of universal gravitation. For a historical account see \cite{Arce,Corda,Ludl}. When estimating these classical tests, it is of great interest to consider as many physical properties of the object that generates the curvature of spacetime (mass, rotation, magnetic field, etc.) as possible, in order to obtain a better estimate of the contribution of each of these properties in the tests. In the following subsections, we obtain mathematical expressions for all the mentioned tests. An expression for the contribution of the magnetic field in terms of the magnetic dipole moment and electric charge is also found.

\subsection{Geodesic equation}

The Lagrangian for both a massive or massless particle is given by the expression

\begin{equation} \label{lagrangian}
    \mathcal{L} = \frac{1}{2} \dot{s}^2= \frac{1}{2} g_{\eta \mu} \dot{x}^\eta \dot{x}^\mu.
\end{equation}

\noindent 
where $ \dot{x^\mu} = {dx^\mu}/{d\sigma} $ ($ x^{\mu} = (t, \, r, \, \theta, \, \varphi) $), $\dot{s} = ds/d\sigma$ and $\sigma$ is an affine parameter.

\noindent 
There are two conserved quantities because the Lagrangian does not depend on $t$ or $\varphi$. These conserved quantities are

\begin{align}
\label{const}
\frac{\partial\mathcal{L}}{\partial \dot{t}} &= -V \dot{t} + W \dot{\varphi} = -E , \\
\frac{\partial\mathcal{L}}{\partial \dot{\varphi}} &=Z \dot{\varphi} + W \dot{t} = L_z , \nonumber
\end{align}

\noindent 
where $E$ and $L_z$ are the energy and angular momentum, respectively. From Eq.(\ref{const}), we get

\begin{align}
\label{velocities}
    \dot{t} &= \frac{EZ+WL_z}{W^2+VZ} \equiv T,\\
    \dot{\varphi} &= \frac{VL_z-EW}{W^2+VZ}  \equiv \Phi. \nonumber
\end{align}

\noindent 
From the lagrangian and Eq.(\ref{velocities}), we have the expression

\begin{align}
\label{drdphi}
\dot{s}^2 = -V T^2 +X \Big(\frac{dr}{d\varphi}\Big)^2\Phi^2 + Y \Big(\frac{d\theta}{d\varphi}\Big)^2\Phi^2 +Z \Phi^2 +2 W T \Phi.
\end{align}

\noindent 
 Solving for $dr / d\varphi$ from Eq.(\ref{drdphi})

\begin{eqnarray}
\Big(\frac{dr}{d\varphi}\Big)^2 = \frac{1}{X\Phi^2}\Big[\dot{s}^2 + VT^2 - Y\Big(\frac{d\theta}{d\varphi}\Big)^2\Phi^2-Z\Phi^2-2WT\Phi\Big] \equiv G\left(r, \theta,\dot{s}\right) . 
\end{eqnarray}

\noindent 
The analysis is made in the equatorial plane $ \theta = {\pi}/{2} $, then

\begin{equation}\label{um}
    \Big(\frac{dr}{d\varphi}\Big)^2= G\left(r, \frac{\pi}{2}, \dot{s} \right)\equiv F(r,\dot{s})= \frac{1}{X\Phi^2}\left[\dot{s}^2+VT^2-Z\Phi^2-2WT\Phi \right].
\end{equation}

\noindent 
Using the change of variable $u = {1} / {r}$, we obtain

\begin{align}
\label{orbit}
    \Big(\frac{du}{d\varphi}\Big)^2 &= u^4F(u,\dot{s}),\\
    \frac{d^2 u}{d\varphi^2} &= 2u^3 F + \frac{1}{2}u^4 \frac{dF}{du}. \nonumber
\end{align}

\noindent 
We used the software Reduce to obtain approximate expressions for classical tests of general relativity by means of Taylor series expansions up to order $ {\cal O}(M^3, \, J^3, \, Q^3, \, \mu^3, \, Q_e^3) $.

\subsection{Angle of the deflection of light}

\noindent 
Since we are studying the photon trajectory, it follows that

\begin{equation}
    \dot{s}^2  = 0 .
\end{equation}

\noindent 
By doing a Taylor expansion of $F(u)$ in Eq.(\ref{orbit}), we have

\begin{align}
\label{difeq}
\frac{d^2 u}{d\varphi^2} + u &= -2b^{-3} J - 8b^{-3} u J M + 12b^4 u J^2 + 2b^{-3} u \mu Q_e + 3u^2 M + 3b^{-2} u^2 Q \nonumber \\
& + 34b^{-2} u^3 J^2 + 10b^{-2} u^3 Q M - 18b^{-3} u^3 Q J - 2u^3 Q_e^2 + \frac{8}{3} b^{-2}u^3 \mu^2  \\
& - \frac{81}{2}u^5 J^2 + \frac{3}{2}u^5 Q M - \frac{93}{4}b^{-2}u^5 Q^2 - \frac{7}{2}u^5 \mu^2 + 33u^7 Q^2 \nonumber \\
&+ {\cal O}(M^3, \, J^3,\, Q^3, \, \mu^3, \, Q_e^3) , \nonumber
\end{align}

\noindent 
where $ b = {L_z}/{E} $ is the impact parameter.

\noindent 
The next step is to find an expression for $b$ in terms of the other constants. Evaluating the equation Eq.(\ref{orbit}) at the maximum value $u_m$ \cite{Bodenner} 

\begin{align}
    \Big(\frac{du}{d\varphi}\Big)^2= u_m^4F(u_m) = 0 ,
\end{align}

\noindent 
where $u_m$ is the inverse of the closest approach in the curved spacetime $r_{min}$, this quantity is illustrated in the Figure \ref{deflection}. By expanding $u_m^4F(u_m)$ and solving for ${1}/{b}$, 
the expression becomes

\begin{align}
\label{close}
    b^{-1} &= u_m - u_m^2 M - \frac{1}{2} u_m^3 M^2 + \frac{1}{2} u_m^3 Q_e^2 - u_m^4 Q - \frac{1}{12} u_m^5 \mu^2 \nonumber \\
&- \frac{7}{4} u_m^5 QM - \frac{7}{4} u_m^5 J^2 + \frac{43}{8} u_m^7 Q^2  + {\cal O}(M^3, \, J^3,\, Q^3, \, \mu^3, \, Q_e^3).
\end{align}

\noindent 
To solve Eq.(\ref{difeq}) a perturbative Ansatz \cite{Arce, Bodenner} is required

\begin{align}
\label{ugeneral}
    u &= u_0 \cos{\varphi} + \sum_{\alpha=1}^{3}\Big[Mu_{M1 \alpha} + Q_eu_{e1 \alpha} \Big] u_0^{(3-\alpha)} u_m^{(\alpha-1)} \nonumber \\
    &+ \sum_{\alpha=1}^{4} \Big[M^2 u_{M2 \alpha}+Q_e^2u_{e2 \alpha} + \mu u_{\mu1\alpha} + J u_{J1\alpha} + MQ_e u_{Me\alpha}\Big]u_0^{(4-\alpha)} u_m^{(\alpha-1)}   \nonumber  \\
    &+ \sum_{\alpha=1}^{5} \Big[Q u_{Q1\alpha} + JM u_{JM\alpha} + JQ_e u_{Je\alpha} + \mu M u_{M\mu \alpha} + Q_e \mu u_{e\mu\alpha}\Big]u_0^{(5-\alpha)} u_m^{(\alpha-1)} \nonumber \\
    &+ \sum_{\alpha=1}^{6} \Big[\mu^2 u_{\mu2\alpha} + J^2 u_{J2\alpha} + Q Q_e u_{Qe\alpha} + J\mu u_{J\mu \alpha} + Q M u_{QM\alpha}\Big]u_0^{(6-\alpha)} u_m^{(\alpha-1)} \\
    &+ \sum_{\alpha=1}^{7}\Big[JQ u_{JQ\alpha} + \mu Q u_{Q\mu \alpha} \Big] u_0^{(7-\alpha)} u_m^{(\alpha-1)}  + \sum_{\alpha=1}^{8} Q^2 u_{Q2\alpha} u_0^{(8-\alpha)} u_m^{(\alpha-1)} , \nonumber
\end{align}

\noindent 
where $u_0$ is the inverse of the closest approach in the flat spacetime $r_0$, this quantity is illustrated in the Figure \ref{deflection}. The other terms are functions that depend on $\varphi$. By substituting $u$ in the differential equation, one can obtain 103 separate differential equations, of which 80 have the trivial solution

\begin{equation}
    y(\varphi) = A \sin{\varphi} + B \cos{\varphi} .
\end{equation}

\noindent 
Because of this, the arbitrary constants will be taken as zero.  
The other differential equations and their solutions are shown in the appendix A. If we substitute the obtained solutions into Eq.(\ref{ugeneral}), $u$ takes the form

\begin{align} \label{u}
    u &= u_0 \cos{\varphi} - \frac{1}{2}QMu_0^3 u_m^2\cos(3\varphi) + Q^2 u_0^7 \mathcal{R}_7  \nonumber \\
    &+ u_0^3\Big( - \frac{4}{3} \mu^2  u_m^2  + Q_e^2   - 17 J^2  u_m^2  + 9JQ u_m^3 \Big) \mathcal{R}_1 \nonumber \\
    &+  u_0 u_m^3 \Big(6 J^2 u_m -10JM  - 6 JQ u_m^2 + \mu Q_e  \Big)\mathcal{R}_2 \nonumber\\
    &+ Ju_m^3\Big(1  - 3 M u_m - 3 Q u_m^3 \Big)\mathcal{R}_3\\
    &+ u_0^5\Big(\frac{7}{27} \mu^2 + \frac{31}{54} Q^2  u_m^2  + J^2  -  \frac{1}{27}QM\Big) \mathcal{R}_4    \nonumber \\
    &+ u_0^2\Big(M  + Q u_m^2  - 2 Q^2  u_m^5  - 2 QM  u_m^3 \Big)\mathcal{R}_5 \nonumber\\
    &+ u_0^3 \Big(M^2   + Q^2  u_m^4  + 3QM u_m^2 \Big) \mathcal{R}_6, \nonumber
\end{align}

\noindent 
where

\begin{align}
\mathcal{R}_1 &= \frac{1}{16} (\cos(3\varphi) - 12 \varphi \sin{\varphi)}, \nonumber \\
\mathcal{R}_2  &= \varphi \sin{\varphi}, \nonumber \\
\mathcal{R}_3  &= -2, \nonumber \\
\mathcal{R}_4  &=  \frac{27}{256}(15 \cos(3\varphi) + \cos(5\varphi) -120 \varphi \sin{\varphi}) ,  \\
\mathcal{R}_5 &= \frac{1}{2}( 3 - \cos(2\varphi)), \nonumber \\
\mathcal{R}_6  &= \frac{3}{16}( 20\varphi \sin{\varphi} +  \cos(3\varphi)), \nonumber \\
\mathcal{R}_7 &= \frac{-11}{1024} (  126 \cos(3\varphi) + 14 \cos(5\varphi) + \cos(7\varphi)-840\varphi \sin{\varphi}). \nonumber 
\end{align}

\noindent 
Evaluating $\varphi=0$ in $u$, solving for $u_0$ (taking up to the second order in $u_0$) and expanding, the result is

\begin{align}
u_0 &= u_m - M u_m^2 + 2 (J + M^2) u_m^3 - (10 J M + Q) u_m^4 + 6 QM u_m^5  \\
&- 10 J Q  u_m^6 + 4 Q^2 u_m^7 + {\cal O}(M^3, \, J^3,\, Q^3, \, \mu^3, \, Q_e^3). \nonumber
\end{align}

\noindent 
We can substitute $\varphi = {\pi}/{2} + \delta $  in Eq.(\ref{u}), see Figure \ref{deflection}. Expanding $u$ in terms of $\delta$ to obtain an expression from which to find the deflection angle, we obtain

\begin{align} \label{equ}
u &= \frac{u_m^2}{256} \Big( 780 J^2 \pi u_m^3 - 1280 J m \pi u_m^2 + 3584 J m u_m^2 - 1632 J \pi Q u_m^4 \nonumber \\
&+ 3584 J Q u_m^4 - 512 J u_m + 480 m^2 \pi u_m 
- 1024 m^2 u_m + 1500 m \pi Q u_m^3\nonumber \\
&- 3072 m Q u_m^3 + 512 m - 12 \mu^2 \pi u_m^3 + 128 \mu \pi Q_e u_m^2 
+ 705 \pi Q^2 u_m^5 \nonumber \\
&- 96 \pi Q_e^2 u_m - 2048 Q^2 u_m^5 + 512 Q u_m^2 \Big) \\
&+ \frac{u_m}{3072} \Big( 21888 J^2 u_m^4 - 3264 J Q u_m^5 - 6144 J u_m^2 + 7104 m^2 u_m^2 \nonumber \\
&+ 17664 m Q u_m^4 + 3072 m u_m + 64 \mu^2 u_m^4 + 3072 \mu Q_e u_m^3 \nonumber \\
&+ 3405 Q^2 u_m^6 + 3072 Q u_m^3 - 1728 Q_e^2 u_m^2 - 3072 \Big)  \delta + {\cal O}(\delta^2) \nonumber .
\end{align}

\noindent 
Now, it applies that if $\delta \approx 0 $, then $ u \approx 0 $, therefore from Eq.(\ref{equ}), we solve for $ \delta $ after doing a Taylor expansion. The result is

\begin{align}\label{edeflection1}
\delta &= 2  M u_m + \Big(\frac{ 15\pi }{8} - 2\Big) M^2 u_m^2 + 2 u_m^3 Q + 3\Big(\frac{705\pi }{768} - 2\Big) Q^2 u_m^6  \nonumber \\
& - 2 J u_m^2 + 2 \Big(\frac{ 195\pi }{128} + 2 \Big) J^2 u_m^4  - \frac{3\pi }{8} u_m^2 Q_e^2 - \frac{3\pi}{64} \mu^2 u_m^4 + \frac{\pi }{2} \mu Q_e u_m^3 \\
&+ 4\Big(\frac{ 375\pi}{256} - 2 \Big)Q M u_m^4   -  4\Big( \frac{5\pi}{4} - 2\Big) JM u_m^3 -4\Big(\frac{ 51\pi }{32} - 2\Big) JQ u_m^5 \nonumber \\
&+ {\cal O}(M^3, \, J^3,\, Q^3, \, \mu^3, \, Q_e^3) . \nonumber
\end{align}

\begin{figure}[H]
    \centering
\begin{tikzpicture}[scale=3.0]
    \draw[fill=yellow] (0,0) circle (0.35) node[above] at (-0.15, 0.0) {$M$};
    \draw[fill=black] (0,0) circle (0.01) node[above] at (-0.18, 0.0) {};
     
      \draw[densely dashed, -] (0.0,0.0) -- (0.0,0.5) node[ above] at (0.13,0.15) {$r_{min}$};

       \draw[densely dashed, -] (0.0,0.0) -- (0.0,-0.4) node[ above] at (0.08,-0.2) {$r_{0}$};
     
    \draw[<-,thick] (-2,-0.4) -- (2,-0.4);
    \draw[-,thick] (0,0.5) -- (-0.5,0.7) node[right] {};
    \draw[-,decorate,decoration={snake,amplitude=1.5mm,segment length=5mm,pre length=1mm,post length=2mm}] (0.0,0.5) -- (2,-0.4);
    \draw[->,decorate,decoration={snake,amplitude=1.5mm,segment length=5mm,pre length=1mm,post length=2mm}] (0.0,0.5) -- (-2,-0.4);
    \draw[->] (1.1,-0.4) +(0:0.24cm) arc (180:105:0.2cm) node[right] at (1.4,-0.3) {$\delta$};
    \draw[->] (-1.5,-0.4) +(0:0.24cm) arc (0:75:0.2cm) node[right] at (-1.6,-0.3) {$\delta$};
    \draw[->] (-0.43,0.67) +(0:0.0cm) arc (155:250:0.17cm) node[right] at (-0.43,0.525) {$\Delta \phi$};
\end{tikzpicture}

\caption{Deflection of light by an object with mass $M$. The quantities $\delta$, $r_0$, $r_{min}$ and $\Delta \phi$ are the partial deflection angle, distance of closest approach in flat spacetime, distance of closest approach in curve spacetime and total deflection angle, respectively.}
    \label{deflection}
\end{figure}

\noindent  
The total deflection angle is $\Delta \phi = 2\delta$, because HT-like metric is symmetric across the polar axis.

\begin{align}
\label{edeflection2}
\Delta \phi &= 4  M u_m+ \Big(\frac{ 15\pi }{4} - 4\Big) M^2 u_m^2 + 4 u_m^3 Q + 3\Big(\frac{705\pi }{384} - 4\Big) Q^2 u_m^6 \nonumber  \\
& - 4 J u_m^2 + 2 \Big(\frac{ 195\pi }{64} + 4 \Big) J^2 u_m^4  - \frac{3\pi }{4} u_m^2 Q_e^2 - \frac{3\pi}{32} \mu^2 u_m^4 + \pi  \mu Q_e u_m^3 \\
&+ 4\Big(\frac{ 375\pi}{128} - 4 \Big)Q M u_m^4   -  4\Big( \frac{5\pi}{2} - 4\Big) JM u_m^3 -4\Big(\frac{ 51\pi }{16} - 4\Big) JQ u_m^5,  \nonumber
\end{align}

\noindent 
The equation Eq.(\ref{edeflection2}) can be written as follows

\begin{align}
\label{ref-def}
\Delta \phi &=  \Delta \phi_{M1} +\Delta \phi_{M2} - \Delta \phi_{Q1} + \Delta \phi_{Q2} - \Delta \phi_{J1} + \Delta \phi_{J2} - \Delta \phi_{Q_e 1} \\
&- \Delta \phi_{\mu 2} + \Delta \phi_{Q_e \mu} - \Delta \phi_{MQ} - \Delta \phi_{JM} +  \Delta \phi_{JQ}.\nonumber
\end{align}

\noindent 
The expression Eq.(\ref{edeflection2}) agrees with Schwarzschild up to second order\cite{Arce,Bodenner}, with Kerr or Lense-Thirring up to second order \cite{Ali,Werner,Paez}, and includes the massive quadrupole, electric charge and magnetic dipole correction up to second order. The first term of Eq.(\ref{ref-def}) correspond to the first term of Eq.(\ref{edeflection2}),  the second term of Eq.(\ref{ref-def}) correspond to the second term of Eq.(\ref{edeflection2}) and so on.  This will help us give a reference of which term is being analyzed.

\subsection{Periastron precession}

\noindent 
For the periastron precession a similar analysis to the one described in the previous subsection is presented, although it is also possible to obtain the periastron precession from the adiabatic theory of motion in any spacetime (See for instance \cite{Boshkayev, Suliyeva}). In this case, we are studying the trajectory of a particle of a finite mass $m$, which means that with respect to the proper time $\tau$, $\dot{s}$ can take the form

\begin{equation}
    \dot{s}^2 = -m^2\Big(\frac{ds}{d\tau}\Big)^2 = -m^2.
\end{equation}

\noindent 
By doing a Taylor expansion of $F(u)$ in Eq.(\ref{orbit}), we obtain

\begin{align}
\label{massive}
\frac{d^2 u}{d\phi^2} + u &= 3 M u^2 + \frac{M}{l_z^2}  - 2\frac{\varepsilon}{l_z^3} (\varepsilon^2 - 1 ) J  - 8\frac{\varepsilon^3}{l_z^3}  J M u \nonumber \\
& + 12\frac{\varepsilon^2}{l_z^4} (\varepsilon^2 - 1)  J^2 u - \frac{Q_e^2}{l_z^2} u  + 2\frac{\varepsilon}{l_z^3} (\varepsilon^2 - 1)  \mu Q_e u  \nonumber \\
& + \frac{3}{2} (2\varepsilon^2 - 1) \frac{Q}{l_z^2}  u^2  + 34 \frac{ J^2}{l_z^2} (\varepsilon^2 - 1) u^3  + 10 \frac{\varepsilon^2}{l_z^2}  Q M u^3  \\
& - 2\frac{\varepsilon}{l_z^3} (9\varepsilon^2 - 5) J Q  u^3  - 2 u^3 Q_e^2 + \frac{8}{3}\frac{\mu^2}{l_z^2} (\varepsilon^2 - 1) u^3  - \frac{81}{2}  J^2 u^5 \nonumber \\
& + \frac{3}{2} Q M u^5  - \frac{3}{4} \frac{Q^2}{l_z^2} (31\varepsilon^2 - 37) u^5  - \frac{7}{2} \mu^2 u^5  + 33 Q^2 u^7 \nonumber\\
&+ {\cal O}(M^3, \, J^3,\, Q^3, \, \mu^3, \, Q_e^3). \nonumber
\end{align}

\noindent 
In Eq.(\ref{massive}), $l_z = L_z/m$  and $\varepsilon = E / m$.  To solve this differential equation, it is better to propose a solution of the form \cite{Arce}

\begin{align}\label{propose}
    u= \frac{1}{l}[1+\omega(\varphi)],
\end{align}

\noindent 
This proposed solution, similar to elliptic orbits, is illustrated in the Figure \ref{precession}, where $l$ is the semi-latus rectum and $\omega(\varphi)$  is a periodic function that satisfies $\omega(\varphi) \ll 1$. Using Eq.(\ref{propose}) in Eq.(\ref{massive}) and expanding in $\omega$, the equation becomes 

\begin{align}
\label{reducida}
 \frac{d^2 \omega}{d\varphi^2} &= \frac{1}{12l_z^4}\Big(144\varepsilon^4J^2 - 96\varepsilon^3Jl_zM - 216\varepsilon^3Jl_z\Big(\frac{Q}{l^2}\Big) - 24\varepsilon^3Jl_z l\nonumber \\
&+ 24\varepsilon^3l_z\mu  Q_e + 408\varepsilon^2J^2\Big(\frac{l_z}{l}\Big)^2 - 144\varepsilon^2J^2 + 120\varepsilon^2l_z^2\Big(\frac{MQ}{l^2}\Big) \nonumber\\
&+ 32\varepsilon^2l_z^2\Big(\frac{\mu}{l}\Big)^2 - 279\varepsilon^2l_z^2\Big(\frac{Q^2}{l^4}\Big)+ 36\varepsilon^2l_z^2\Big(\frac{Q}{l}\Big) + 120 \varepsilon Jl_z\Big(\frac{Q}{l^2}\Big) \nonumber \\
&+ 24\varepsilon Jl_zl - 24\varepsilon l_z\mu   Q_e - 486J^2\Big(\frac{l_z}{l}\Big)^4 - 408J^2\Big(\frac{l_z}{l}\Big)^2 + 18MQ\Big(\frac{l_z}{l}\Big)^4 \nonumber \\
&+ 36l_z^4\Big(\frac{M}{l}\Big) - 42l_z^4 \Big(\frac{\mu^2}{l^4}\Big) + 396l_z^4\Big(\frac{Q^2}{l^6}\Big) - 24l_z^4\Big(\frac{Q_e}{l}\Big)^2 - 12l_z^4 \nonumber \\
&+ 12l_z^2 Ml - 32l_z^2 \Big(\frac{\mu}{l}\Big)^2 + 333l_z^2\Big(\frac{Q^2}{l^4}\Big) - 18l_z^2\Big(\frac{Q}{l}\Big) - 12l_z^2Q_e^2\Big) \\
& + \frac{1}{4l_z^4} \Big(48\varepsilon^4J^2 - 32\varepsilon^3Jl_z M - 216\varepsilon^3Jl_z\Big(\frac{Q}{l^2}\Big) + 8\varepsilon^3l_z\mu  Q_e \nonumber \\
&+ 408\varepsilon^2l_z^2\Big(\frac{J}{l}\Big)^2 - 48\varepsilon^2J^2 + 120\varepsilon^2l_z^2\Big(\frac{MQ}{l^2}\Big) + 32\varepsilon^2l_z^2 \Big(\frac{\mu}{l}\Big)^2 \nonumber \\
&- 465\varepsilon^2l_z^2\Big(\frac{Q^2}{l^4}\Big) + 24\varepsilon^2l_z^2\Big(\frac{Q}{l}\Big) + 120 \varepsilon Jl_z\Big(\frac{Q}{l^2}\Big) - 8 \varepsilon l_z\mu  Q_e \nonumber \\
&- 810J^2\Big(\frac{l_z}{l}\Big)^4 - 408J^2\Big(\frac{l_z}{l}\Big)^2 + 30l_z^4\Big(\frac{MQ}{l^4}\Big) + 24l_z^4\Big(\frac{M}{l}\Big) \nonumber \\
&- 70l_z^4 \Big(\frac{\mu^2}{l^4}\Big) + 924l_z^4\Big(\frac{Q^2}{l^6}\Big) - 24l_z^4\Big(\frac{Q_e}{l}\Big)^2 - 4l_z^4 - 32l_z^2 \Big(\frac{\mu}{l}\Big)^2 \nonumber \\
&+ 555l_z^2\Big(\frac{Q}{l^4}\Big) - 12l_z^2\Big(\frac{Q}{l}\Big) - 4l_z^2Q_e^2\Big) \omega + {\cal O}(\omega^2).  \nonumber
\end{align}

\noindent 
From here, it is clear that the solution takes the form 

\begin{align}
    \omega(\varphi)= A\sin\left(\sqrt{\Omega}\varphi\right)+B\cos\left(\sqrt{\Omega}\varphi\right),
\end{align}

\begin{figure}[H]
    \centering
\begin{tikzpicture}[scale=1.91]

    \draw[fill=yellow] (0,0) circle (0.4) node[right] at (-0.15,-0.2) {$M$};

    \draw[thick] (-1.3228756555322954,0) ellipse (2 and 1.5);
    \draw[densely dashed, -] (-1.3228756555322952,0) -- (0.6771243444677,0) node[above] at (-0.5,0.0) {$a$};
    \draw[densely dashed, -] (-1.3228756555322952,0) -- (-1.3228756555322952,-1.5) node[midway, right] {$b$};
    \draw[densely dashed, -] (0,0) -- (0,1.125) node[midway, right] {$l$};
    \draw[thick, rotate around={45:(0,0)}] (-1.3228756555322954, 0) ellipse (2 and 1.5);
     \draw[densely dashed, -] (-0.9354143566934853463959,-0.9354143566934853463959) -- (0.4787992156548067,0.4787992156548067) node[ above] at (-0.39,-0.38) {$a$};
    \draw[densely dashed, -] (-0.9354143566934853463959,-0.9354143566934853463959) -- (0.12524581508633694020,-1.996074528473306632997) node[ right] at (-0.5,-1.2)  {$b$};
    \draw[densely dashed, -] (0,0) -- (-0.7954951288348661,0.7954951288348661) node[midway, right] {$l$};
    
    \filldraw[gray] (0.6771243444677,0) circle (0.04) node[above right] {\textcolor{black}{}};
    \filldraw[gray] (0.4787992156548067,0.4787992156548067) circle (0.04) node[above right] {\textcolor{black}{}};
    \draw[->] (0.45,0.0) +(0:-0.15cm) arc (0:45:0.3cm) node[right] at (0.65,0.2) {$\Delta \varphi$};
   
\end{tikzpicture}
\caption{Periastron precession due to the curvature of spacetime, the lighter object is shown with a small filled circles which is orbiting around a heavier object of mass M. The quantities $l$, $\Delta \varphi$, $a$ and $b$ are semi-latus rectum, periastron precession,  semi-major and semi-minor axis, respectively. } 
    \label{precession}
\end{figure}

\noindent 
where

\begin{align}\label{omega}
\Omega &= -\frac{1}{4l_z^4} \Big(48\varepsilon^4J^2 - 32\varepsilon^3Jl_z M - 216\varepsilon^3Jl_z\Big(\frac{Q}{l^2}\Big) + 8\varepsilon^3l_z\mu  Q_e \nonumber \\
&+ 408\varepsilon^2l_z^2\Big(\frac{J}{l}\Big)^2 - 48\varepsilon^2J^2 + 120\varepsilon^2l_z^2\Big(\frac{MQ}{l^2}\Big) + 32\varepsilon^2l_z^2 \Big(\frac{\mu}{l}\Big)^2 \nonumber \\
&- 465\varepsilon^2l_z^2\Big(\frac{Q^2}{l^4}\Big) + 24\varepsilon^2l_z^2\Big(\frac{Q}{l}\Big) + 120 \varepsilon Jl_z\Big(\frac{Q}{l^2}\Big) - 8 \varepsilon l_z\mu  Q_e  \\
&- 810J^2\Big(\frac{l_z}{l}\Big)^4 - 408J^2\Big(\frac{l_z}{l}\Big)^2 + 30l_z^4\Big(\frac{MQ}{l^4}\Big) + 24l_z^4\Big(\frac{M}{l}\Big) \nonumber \\
&- 70l_z^4 \Big(\frac{\mu^2}{l^4}\Big) + 924l_z^4\Big(\frac{Q^2}{l^6}\Big) - 24l_z^4\Big(\frac{Q_e}{l}\Big)^2 - 4l_z^4 - 32l_z^2 \Big(\frac{\mu}{l}\Big)^2 \nonumber \\
&+ 555l_z^2\Big(\frac{Q}{l^4}\Big) - 12l_z^2\Big(\frac{Q}{l}\Big) - 4l_z^2Q_e^2\Big)  \nonumber.
\end{align}

\noindent 
A cycle is completed if

\begin{align}
\label{phi}
    \varphi= \frac{2\pi}{\sqrt{\Omega}},
\end{align}

\noindent 
If expand Eq.(\ref{phi}), we obtain

\begin{align} \label{fullexp}
\varphi &= 2\pi + 6\pi  \left(\frac{M}{l}\right) + 27\pi  \left(\frac{M}{l}\right)^2 - 8\pi\frac{\varepsilon^3}{l_z^3} J M + 12 \pi \frac{\varepsilon^2}{l_z^4} (\varepsilon^2 - 1) J^2 \nonumber \\
& + 102 \frac{\pi }{l_z^2} (\varepsilon^2 - 1) \left(\frac{J}{l}\right)^2 - \frac{405\pi}{2}  \left(\frac{J^2}{l^4}\right) + \frac{8\pi}{l_z^2} (\varepsilon^2 - 1) \left(\frac{\mu}{l}\right)^2 \nonumber \\
&- \frac{35\pi}{2}  \left(\frac{\mu^2}{l^4}\right) + \frac{3\pi }{l_z^2} (2\varepsilon^2 - 1)  \left(\frac{Q}{l}\right) + \frac{3\pi }{l_z^2} (28\varepsilon^2 - 9)  \left(\frac{Q M}{l^2}\right)  \\
& + \frac{15\pi}{2}  \left(\frac{Q M}{l^4}\right) - 6\pi \frac{\varepsilon }{l_z^3} (9\varepsilon^2 - 5)   \left(\frac{JQ}{l^2}\right) + \frac{27\pi }{4l_z^4} (2\varepsilon^2  + 1)^2 \left(\frac{Q}{l}\right)^2 \nonumber \\
& - \frac{15\pi }{4l_z^2} (31\varepsilon^2 - 37)  \left(\frac{Q^2}{l^4}\right) + 231\pi  \left(\frac{Q^2}{l^6}\right) + 2\pi\frac{\varepsilon }{l_z^3} (\varepsilon^2 - 1) Q_e \mu \nonumber \\
& - \frac{ \pi}{l_z^2 l^2} (l^2 +  6l_z^2) Q_e^2  + {\cal O}(M^3, \, J^3,\, Q^3, \, \mu^3, \, Q_e^3).  \nonumber
\end{align}

\noindent 
The next step is to find relation for $\varepsilon$ in terms of the semi-latus rectum and $l_z$ for obtain gravitational potential (See Appendix B). This can be found with the post-newtonian relation , following the procedure shown in \cite{Subra}.

\begin{align}\label{E}
\varepsilon &= -1 + \left(\frac{M}{l}\right)  - \frac{1}{2} \left(\frac{l_z}{l}\right)^2 + \frac{1}{2} \left(\frac{M}{l}\right)^2  - \frac{1}{2} \left(\frac{Q_e}{l}\right)^2 \nonumber \\
& + \frac{l_z^2}{2} \left(\frac{M}{l^3}\right)  + \frac{1}{2} \left(\frac{Q}{l^3}\right)  + 2 l_z \left(\frac{J}{l^3}\right)  + \frac{1}{8} \left(\frac{l_z}{l}\right)^4 \nonumber \\
& + \frac{l_z^2}{4} \left(\frac{M^2}{l^4}\right)  + \left(\frac{MQ}{l^4}\right)  - 2 \left(\frac{J^2}{l^4}\right) - \frac{l_z^2}{4} \left(\frac{Q_e^2}{l^4}\right)  \\
& - l_z \left(\frac{\mu Q_e}{l^4}\right)  - \frac{l_z^4}{8} \left(\frac{M}{l^5}\right) + \frac{3 l_z^2}{4} \left(\frac{Q}{l^5}\right) - \frac{1}{16} \left(\frac{l_z}{l}\right)^6 \nonumber \\
& - \frac{l_z^4}{16} \left(\frac{M^2}{l^6}\right)  + \frac{5 l_z^2}{4} \left(\frac{QM}{l^6}\right)  - \frac{1}{8} \left(\frac{Q^2}{l^6}\right)  - \frac{7 l_z}{2} \left(\frac{J Q}{l^6}\right)  \nonumber \\
& - \frac{5 l_z^2}{4} \left(\frac{J^2}{l^6}\right) + \frac{l_z^4}{16} \left(\frac{Q_e^2}{l^6}\right)  + \frac{l_z^2}{12} \left(\frac{\mu^2}{l^6}\right) \nonumber\\
&+ {\cal O}(M^3, \, J^3,\, Q^3, \, \mu^3, \, Q_e^3) \nonumber.
\end{align}

\noindent 
Using Eq.(\ref{E}) in Eq.(\ref{fullexp}), the angle after expanding becomes 

\begin{align}
\varphi &= 2\pi + \Delta \varphi,
\end{align}

\noindent 
From which the periastron precession, $\Delta \varphi$ is found to be 

\begin{align}
\label{prece}
\Delta \varphi &= 6\pi \left(\frac{M}{l}\right) + 27\pi \left(\frac{M}{l}\right)^2  - \frac{19\pi }{2} \left(\frac{\mu^2}{l^4}\right)  \nonumber \\
&+  \frac{\pi}{4l_z} \left[ \left(\frac{l_z}{l}\right)^4 -  \left(\frac{2l_z}{l}\right)^2 - 8\right] \left(\frac{Q_e \mu}{l^2}\right)  \nonumber \\
& - \frac{\pi}{l_z^2} \left[6 \left(\frac{l_z}{l}\right)^2 + 1\right] Q_e^2 + \frac{3\pi}{l_z^2}  \left[2 \left(\frac{l_z}{l}\right)^2 + 1\right] \left(\frac{Q}{l}\right) \nonumber \\
&+ \frac{3\pi}{2l_z^2} \left[53 \left(\frac{l_z}{l}\right)^2 + 30\right] \left(\frac{Q M}{l^2}\right) \\
&+ \frac{3\pi}{4l_z^4} \left[173 \left(\frac{l_z}{l}\right)^4 + 58\left(\frac{l_z}{l}\right)^2 + 9\right] \left(\frac{Q}{l}\right)^2 \nonumber \\
&+ \frac{\pi}{2l_z^3} \left[- \left(\frac{l_z}{l}\right)^6 + 6\left(\frac{l_z}{l}\right)^4 + 24 \left(\frac{l_z}{l}\right)^2 + 16\right] J M \nonumber \\
& + \frac{6\pi}{l_z^3}   \left[2 \left(\frac{l_z}{l}\right)^4 + 7 \left(\frac{l_z}{l}\right)^2 + 4\right] \left(\frac{J Q}{l^2}\right) \nonumber \\
& + \frac{3\pi}{2l_z^2}  \left[-59 \left(\frac{l_z}{l}\right)^2 + 8\right] \left(\frac{J}{l}\right)^2 \nonumber.
\end{align}

\noindent 
The last equation can be transformed into

\begin{align}
\label{prece3}
\Delta \varphi &=  \Delta \varphi_{M1} + \Delta \varphi_{M2} - \Delta \varphi_{\mu 2} + \Delta \varphi_{Q_e \mu} - \Delta \varphi_{Q_e 2} - \Delta \varphi_{Q1} \\
&- \Delta \varphi_{QM} + \Delta \varphi_{Q2} + \Delta \varphi_{JM} - \Delta \varphi_{JQ} + \Delta \varphi_{J2} \nonumber. 
\end{align}

\noindent 
The first term of Eq.(\ref{prece}) correspond to the first term of Eq.(\ref{prece3}),  the second term of Eq.(\ref{prece}) correspond to the second term of Eq.(\ref{prece3}) and so on. The equation Eq.(\ref{prece}) reduce to Schwarzschild periastron precession  when $J$, $Q$, $\mu$ and $Q_e$ are zero. This result is in agreement with \cite{Subra} in angular momentum contribution.

\subsection{Time delay}

\noindent The time delay was proposed by I.I. Shapiro in 1964 \cite{Shapiro}. This test is considered the fourth test, comparing  travel time in curved spacetime with that in flat spacetime. To start, let us take $\theta={\pi}/{2}$, the first term of Eq.(\ref{u}) and its derivative \cite{Dinverno},

\begin{eqnarray}\label{trigonometric}
   r\cos{\varphi}=r_0,&\nonumber\\
   \frac{dr}{d\varphi}\cos{\varphi}- r \sin{\varphi}= 0, \nonumber&\\
    \frac{d\varphi}{dr}= \frac{1}{r} \cot{\varphi} = \frac{r_0}{r\sqrt{r^2-r_0^2}}  &\\
    \frac{dr}{du} = -\frac{1}{u^2} \nonumber &\\
    d\varphi = -  \frac{du}{\sqrt{u_0^2-u^2}}. \nonumber
\end{eqnarray}

\noindent 
The quantities in these trigonometric relations are represented in Figure \ref{time-delay}.

\begin{figure}[H]
    \centering
\begin{tikzpicture}[scale=3.0]
    \draw[fill=yellow] (0,0) circle (0.3) node[above] at (0,-0.2)  {$M$};
    \draw[fill=gray] (-1.5,0.4) circle (0.05) node[above] at (-1.75,0.4)  {$Source$};
    \draw[fill=gray] (1.5,0.4) circle (0.05) node[above] at (1.8,0.4)  {$Observer$};
    \draw[->,thick] (-1.5,0.4) -- (0,0.4) node[right] {};
    \draw[-,thick] (0,0.4) -- (1.5,0.4) node[right] {};
    \draw[-, thick] (-1.5,0.4) -- (0,0) node[right] at (-0.8,0.1) {$D_1$};
    \draw[-, thick] (0,0) -- (1.5,0.4) node[right] at (0.8,0.1) {$D_2$};
    \draw[densely dashed, -] (0,0) -- (0,0.4) node[above] at (0.08,0.12) {$r_0$};
    
    \draw[->] (0,0.2) +(0:0.0cm) arc (90:160:0.2cm) node[right] at (-0.15,0.1) {$\varphi$};
\end{tikzpicture}
    \caption{Time delay observed in a gravitating object near $M$. The quantities $r_0$, $D_1$ and $D_2$ are  distance of closest approach in flat spacetime, distance between $M$ and Source and distance between $M$ and Observer, respectively.   }
    \label{time-delay}
\end{figure}

\noindent 
Replacing Eq.(\ref{trigonometric}) in Eq.(\ref{LWP})

\begin{align}
d s^2 =-V d t^2 +\frac{X}{u^4} d u^2 + \frac{Z}{(u_0^2-u^2)}du^2 - \frac{2W}{\sqrt{u_0^2-u^2}} d t d u.
\end{align}

\noindent 
For photon, we have $ds=0$, so that

\begin{align}\label{tquadratic}
0=-V \left(\frac{dt}{du}\right)^2 + \frac{X}{u^4}  + \frac{Z}{(u_0^2-u^2)} - \frac{2W}{\sqrt{u_0^2-u^2}} \left(\frac{dt}{du}\right) .
\end{align}

\noindent 
Solving Eq.(\ref{tquadratic}) for ${dt}/{du}$, taking the physical solution, expanding and returning to the variable r to integrate, it is obtained that

\begin{align}
\label{differential}
dt &= \frac{  dr}{\sqrt{r^2 - r_0^2}} \Big(r + 2M - \frac{Mr_0^2}{r^2}  + \frac{4 M^2}{r}   - \frac{2M^2r_0^2}{r^3} - \frac{M^2r_0^4}{2r^5}  - \frac{5J^2}{r^3}\nonumber \\
&+ \frac{27 J^2 r_0^2}{4r^5}  - \frac{4 J Mr_0}{r^3}  
- \frac{ J Q r_0  }{2r^5}   - \frac{2Jr_0}{r^2}  +  \frac{5MQ}{ r^3}  - \frac{5 M Q r_0^2}{4r^5}   + \frac{31 Q^2 }{8r^5}\\
& - \frac{33 Q^2 r_0^2 }{8r^7}  + \frac{Q}{r^2} - \frac{\mu^2 }{2r^3}   + \frac{7 \mu^2 r_0^2  }{12r^5}  + \frac{\mu Q_e r_0}{r^3}  - \frac{Q_e^2}{r}  + \frac{ Q_e^2 r_0^2}{2r^3}  \Big)\nonumber \\
&+ {\cal O}(M^3, \, J^3,\, Q^3, \, \mu^3, \, Q_e^3). \nonumber
\end{align}

\noindent 
The equation Eq.(\ref{differential}) is according Schwarzschild, Kerr, HT \cite{Arce,Dinverno} and contain electric charge and magnetic dipole contributions \cite{Arce,Dinverno}. Integrating this expression, we find that

\begin{align}
\label{tt}
t = l_2 + l_1 + \Delta t. 
\end{align}

\noindent 
The first and two terms in Eq.(\ref{tt}) are the time travel in flat spacetime, thus the time delay is given by

\begin{align}
\label{dt}
\Delta t &= \Big( 2 \ln\left(\Omega_2 \Omega_1\right)-\Gamma_1\Big) M +\frac{1}{16}\Big(45\frac{\Lambda}{r_0} -19 \Gamma_2 - 2r_0^2\Gamma_4 \Big) M^2 \nonumber \\
&- \frac{1}{96}\Big(3\frac{\Lambda}{r_0^3}  +3\frac{\Gamma_2}{r_0^2} - 14 \Gamma_4\Big) \mu^2 - 2 \frac{\Gamma_1}{r_0} J - 2\Big(\frac{\Lambda}{r_0^2}+\frac{\Gamma_2}{r_0}\Big) JM + \frac{\Gamma_1}{r_0^2} Q\nonumber \\
&+\frac{1}{32} \Big(\frac{\Lambda}{r_0^3} + \frac{\Gamma_2}{r_0^2} + 54\Gamma_4   \Big) J^2 + \frac{1}{2}\Big( \frac{\Lambda}{r_0^2} + \frac{\Gamma_2}{r_0}  \Big) \mu Q_e - \frac{1}{4}\Big(3\frac{\Lambda}{r_0} - \Gamma_2 \Big)Q_e^2  \\
& + \frac{5}{32} \Big(13\frac{\Lambda}{r_0^3} + 13 \frac{\Gamma_2}{r_0^2}- 2 \Gamma_4 \Big) QM -\frac{1}{16}\Big(3\frac{\Lambda}{r_0^4} + 3 \frac{\Gamma_2}{r_0^3} + 2\frac{\Gamma_4}{r_0} \Big) Q J \nonumber \\
&+ \frac{1}{128}\Big( 21\frac{\Lambda}{r_0^5} + 21 \frac{\Gamma_2}{r_0^4} + 14 \frac{\Gamma_4}{r_0^2} -88\Gamma_6  \Big) Q^2 , \nonumber
\end{align}

\noindent
where 

\begin{eqnarray}
    &l_n =\sqrt{D_n^2 - r_0^2}\nonumber,\\ 
    &\Gamma_{n} = \frac{l_2 D_1^n + l_1 D_2^n}{ D_2^n D_1^n} ,  \\
    &\Omega_{n} = \frac{1}{r_0} (l_n + D_n) \nonumber,\\
    &\Lambda= 2\arctan(\Omega_1) +  2\arctan(\Omega_2) - \pi . \nonumber
\end{eqnarray}

\noindent 
The last equation can be written as follows 

\begin{align}
\label{dt2}
\Delta t &= \Delta t_{M1} + \Delta t_{M2} - \Delta t_{\mu 2} - \Delta t_{J1} - \Delta t_{JM} - \Delta t_{Q1} + \Delta t_{J2}\\
& + \Delta t_{Q_e \mu} - \Delta t_{Q_e 2} - \Delta t_{MQ} + \Delta t_{JQ} + \Delta t_{Q2}. \nonumber
\end{align}

The first term of Eq.(\ref{dt}) correspond to the fist term of Eq.(\ref{dt2}),  the second term of Eq.(\ref{dt}) correspond to the second term of Eq.(\ref{dt2}) and so on.

\subsection{Gravitational redshift}

To calculate the redshift factor, a comparison of the proper time is performed for observers located at two different distance values. 
In the equatorial plane, $\theta = \pi/2$. This quantity is given by \cite{ Arce, Dinverno}

\begin{align*}
z = &\frac{\lambda - \lambda_0}{\lambda_0}=\sqrt{\frac{g_{tt}(r)}{g_{tt}(r_0)}} - 1 ,
\end{align*}

\noindent 
where $g_{tt} = - V$. 
\\

\noindent Using a Taylor expansion, we obtain

\begin{align}
\label{red}
z &=  \left(\frac{1}{r_0} - \frac{1}{r} \right) M + \frac{1}{2}\left(\frac{3}{r_0^2} - \frac{2}{r_0 r} - \frac{1}{r}\right) M^2 + \frac{1}{2}\left(\frac{1}{r_0^3} - \frac{1}{r^3}\right) Q \nonumber \\
& + \frac{1}{2}\left(\frac{4}{r_0^4} - \frac{1}{r_0 r^3} - \frac{1}{r_0^3 r} - \frac{2}{r^{ 4}}\right) Q M + \frac{1}{8}\left(\frac{1}{r_0^6} - \frac{2}{r_0^3 r^{3}} + \frac{1}{r^{6}}\right) Q^2 \\
&- \frac{1}{2}\left(\frac{1}{r_0^2} - \frac{1}{r^{2}}\right) Q_e^2  + {\cal O}(M^3, \, J^3,\, Q^3, \, \mu^3, \, Q_e^3). \nonumber
\end{align}

\noindent
Considering the first distance on the surface of a compact object ($r_0$) and the second one at  infinity ($r \rightarrow \infty$), we have

\begin{align}
\label{red2}
z_\infty &= \frac{M}{r_0} + \frac{3M^2}{2r_0^2} + \frac{Q}{2r_0^3} + \frac{2 Q M}{r_0^4} 
+ \frac{Q^2}{8r_0^6} - \frac{Q_e^2}{2r_0^2}.
\end{align}

\noindent The Eq.(\ref{red2}) can be written as follows

\begin{align}
\label{red3}
z_\infty &= z_{M1} + z_{M2} - z_{Q1} - z_{MQ} + z_{Q2} - z_{Q_e 1}. 
\end{align}

\noindent The first term of Eq.(\ref{red2}) correspond to the fist term of Eq.(\ref{red3}),  the second term of Eq.(\ref{red2}) correspond to the second term of Eq.(\ref{red3}) and so on.

\section{Numerical Estimations}
\label{num-est}

In this section, we describe the calculation for the magnetic field and the estimation of the magnetic parameter $\mu$ for various objects. With this information, one can provide numerical estimates for the different classical tests for realistic objects. Previous studies have estimated deflection angles using either the McGill Magnetar Catalog or by assuming a magnetar with surface magnetic field of $B_{S} = 10^{11} \, {\rm T}$ \cite{Beissen, Kim}.  Both works assume a mass of $1.4 \, M_{\odot}$ and a radius of $10 \, {\rm km}$. Since the exact masses and radii of these systems are difficult to determine, and their dependence on the equation of state (EoS) is unclear, we adopt the same values for both, mass and radius.

\begin{table}[h]
\centering
\caption{Numerical estimates for the mass, angular momentum and massive quadrupole contributions to the angle of light deflection, periastron precession, time delay and gravitational redshift for different magnetars and magnetar candidates of the McGill Magnetar Catalog, taking $M = 1.4 \, M_{\odot} , \, J/M = 1 \, {\rm m}, \, Q = -9648\, {\rm m^3}, \, Q_e = 0 \, {\rm m} $ and $r_{min}= r_0 = R=10 \, {\rm km}$. \cite{OK}}%
\begin{tabular}{@{}lllll@{}}
\toprule
& M $1^{\text{st}}$ order & M $2^{\text{nd}}$ order & J $1^{\text{st}}$ order   \\
\midrule
$\Delta \phi$ (as) 
& 1.7061$\times10^5$       & 6.8630$\times10^4$        &  1.7061$\times10^1$     \\
$\Delta \varphi$ (as/cent) & 6.0129$\times10^1$        & 1.0089$\times10^{-5}$     & \hspace{0.5cm} -    \\
$\Delta t$ (s)             & 9.6503 $\times 10^{-4}$    & 1.2603 $\times 10^{-5}$    &  2.7591 $\times 10^{-9}$     \\
z  &$2.0679\times10^{-1}$        & $6.4142\times10^{-2}$     & \hspace{0.5cm} -    \\
\bottomrule
\end{tabular}

\begin{tabular}{@{}lllll@{}}
\toprule
& J $2^{\text{nd}}$ order & Q $1^{\text{st}}$ order &  Q $2^{\text{nd}}$ order  \\
\midrule
$\Delta \phi$ (as) 
& 2.3942$\times10^{-3}$       &   7.9602$\times10^{-3}$        & 1.0182$\times10^{-10}$    \\
$\Delta \varphi$ (as/cent) & 5.3678$\times10^{-20}$        &  1.6792$\times 10^{-12}$     & 7.8682$\times10^{-33}$    \\
$\Delta t$ (s)             & 1.4003 $\times 10^{-15}$    &  6.4365 $\times 10^{-13}$    & 1.6003 $\times 10^{-21}$     \\
z  &  \hspace{0.5cm} -        & $4.8240\times10^{-9}$     & $1.1636\times10^{-17}$   \\
\bottomrule
\end{tabular}

\begin{tabular}{@{}lllll@{}}
\toprule
& JM $1^{\text{st}}$ order & JQ $1^{\text{st}}$ order &  MQ $1^{\text{st}}$ order  \\
\midrule
$\Delta \phi$ (as) 
&  1.3597$\times10^{1}$       & 9.8992$\times10^{-7}$   & 8.5660$\times10^{-3}$     \\
$\Delta \varphi$ (as/cent) & 1.2043 $\times10^{-5}$        & 3.0394$\times 10^{-15}$     &  9.3915 $\times 10^{-19}$    \\
$\Delta t$ (s)             &  8.9622 $\times 10^{-10}$    & 3.9201 $\times 10^{-18}$    &  4.2468 $\times 10^{-13}$     \\
z  &\hspace{0.5cm} -        & \hspace{0.5cm} -     & $3.9902\times10^{-9}$   \\
\bottomrule
\end{tabular}
\label{inv}
\end{table}

\noindent
Regarding the rotation of the systems, while the periods of the systems are known, the exact angular momentum would also depend on the EoS. To address this, we calculate an upper bound for the angular momentum consistent with the slow-rotation approximation. This allows us to compare the magnetic dipole contribution to the classical tests with the rotational contribution without concerns about a larger rotation potentially altering the comparison. The smallest period of the McGill Magnetar Catalog is $ \sim 1.36 \, {\rm s} $. Hence, regardless of the mass distribution of a specific magnetar, its angular momentum must be less than that of a spherical shell of the same mass. Thus, we have:

\begin{equation}
    J<\left(\frac{2}{3}MR^2 \right)\left( \frac{2\pi}{T}\right).
    \label{Jmin}
\end{equation}

\noindent 
Taking this into account, we assume a spin parameter of $J/M = 1 \, {\rm m}$. Although NSs are often associated with rapid rotation, magnetars tend to have much longer periods compared to normal radio pulsars \cite{Jawor}. The chosen value of $J/M$ is sufficiently large to allow for a meaningful comparison between the rotational and magnetic dipole contributions to the classical tests, while still being small enough for validity within the aforementioned approximation.

\noindent
As for the quadrupole moment, we use the result found in Laarakkers and Poisson \cite{Poisson}, who relate the angular momentum to the quadrupole moment as follows:

\begin{equation}
    Q \simeq -\alpha \frac{J^2}{M} ,
    \label{QJ}
\end{equation}

\noindent 
where $\alpha$ depends on the EoS and the mass of the object. Laarakkers and Poisson also provide values for $\alpha$ based on these parameters. For a system with $1.4 \, M_{\odot}$, we average the values for $\alpha$ for each EoS, yielding $Q \simeq -9648 \, {\rm m}^3$ \cite{Poisson}.\footnote[1]{For clarity, in SI units a factor of $c$ is added to the denominator, so that $J_{\text{SI}}/(M_{\text{SI}}c) = 1 \, {\rm m} $ and $Q_{\text{SI}} \simeq - ({\alpha}/{c^2}) ({J_{\text{SI}}^2}/{M_{\text{SI}}})$.}


\noindent 
The primary focus of this research is to analyze the contribution of the magnetic dipole when considering the detailed treatment of the magnetic field and the spacetime curvature it produces. A more accurate treatment of the magnetic field is necessary for estimating deflection angles and other general relativity tests.

\noindent 
Even though the calculations assume a magnetic dipole, it is essential not only to apply the approximate equation for a magnetic dipole but also to account for how the magnetic field behaves within the given metric. The norm of the magnetic field is expressed as:

\begin{align}
\begin{split}
    H^2 &= g^{rr}H_{r}^2+g^{\theta \theta} H_{\theta}^2\\
    &= \frac{1}{XYZ}\Big[\Big(\partial_\theta A_\varphi\Big)^2 + \Big(\partial_r A_\varphi\Big)^2 \Big] .
\end{split}
\end{align}

\noindent 
This formulation allows us to compute the magnetic field in any region of space. Since deriving an analytic expression for $\mu$ as a function of $B$ ($H$) is difficult, we calculate $\mu$ computationally for specific sets of values $(B_{s}, r, \theta)$. For $B_{s}, r = 10 , {\rm km}$, and $\theta = \pi$, we follow the methodology outlined by Olausen and Kaspi to determine the surface dipolar magnetic field strength of magnetars as listed in the McGill Magnetar Catalog, where $B_s$ is measured at the poles \cite{JJ,KUY,OK}.

\noindent 
Although the motivation for using magnetars is the strong magnetic field, this can reduce the possible contributions by the rotation and quadrupole moment. To address this, we perform additional calculations for the system PSR B1257+12, which has an orbital period of approximately 6,2185 ${\rm ms}$ \cite{Konacki}. This millisecond pulsar is of particular interest because exoplanets have been discovered orbiting it, enabling calculations for phenomena such as periastron precession and time delay \cite{Wolszczan}.

\noindent 
To estimate the surface magnetic field, we use the well known relation \cite{Handbook}   
\begin{equation}
    B_s = 3.2\times10^{19} \text{ G} \sqrt{P\dot{P}},
\end{equation}

\noindent 
and find a value of $B_s \approx 8.5326\times 10^{8}$ G. The value $\dot{P} = 1.14334\times 10^{-19}$ is also found in \cite{Konacki}. The same analysis as in Eq. (\ref{Jmin}) and Eq. (\ref{QJ}) is done to estimate the values of $J$ and $Q$. In this occasion, after doing the analysis we take $J/M = 225$ m. The mass and radius of the pulsar are taken to be $M = 1.4M_{\odot}$ and $r=10$ km as well.

\begin{table}[h]
\centering
\caption{Numerical estimates for the angle of light deflection for different magnetars and magnetar candidates of the McGill Magnetar Catalog, taking $M = 1.4 \, M_{\odot} , \, J/M = 1 \, {\rm m}, \, Q = -9648\, {\rm m^3}, \, Q_e = 0 \, {\rm m} $ and $r_{min}= r_0 = R=10 \, {\rm km}$. \cite{OK}}%
\begin{tabular}{@{}lllll@{}}
\toprule
Magnetar              & \multicolumn{1}{c}{\begin{tabular}[c]{@{}c@{}}Magnetic field \\ $10^9$  \rm T\end{tabular}} & \multicolumn{1}{c}{\begin{tabular}[c]{@{}c@{}}$\mu$ \\ $10^1$ $\rm m^2$ \end{tabular}} & \multicolumn{1}{c}{\begin{tabular}[c]{@{}c@{}}$ \Delta \phi $\\ $10^5$ as\end{tabular}} & \multicolumn{1}{c}{\begin{tabular}[c]{@{}c@{}}$\Delta \phi_\mu$ \\ $10^{-8}$ as\end{tabular}} \\ \midrule
CXOU J010043.1-721134 & 39.3                                                                                        & 158.58                                                                                                                    & 2.3921                                                                                       & 1527.6                                                                                            \\
4U 0142+61            & 13.4                                                                                        & 54.069                                                                                                                    & 2.3921                                                                                       & 177.60                                                                                            \\
SGR 0418+5729         & 0.61                                                                                        & 2.4613                                                                                                                    & 2.3921                                                                                       & 0.3680                                                                                            \\
SGR 0501+4516         & 18.7                                                                                        & 75.455                                                                                                                    & 2.3921                                                                                       & 345.87                                                                                            \\
SGR 0526-66           & 56.0                                                                                        & 225.96                                                                                                                    & 2.3921                                                                                       & 3101.8                                                                                            \\
1E 1048.1-5937        & 38.6                                                                                        & 155.75                                                                                                                    & 2.3921                                                                                       & 1473.7                                                                                            \\
1E 1547.0-5408        & 31.8                                                                                        & 128.31                                                                                                                    & 2.3921                                                                                       & 1000.2                                                                                            \\
PSR J1622-4950        & 27.4                                                                                        & 110.56                                                                                                                    & 2.3921                                                                                       & 742.57                                                                                            \\
SGR 1627-41           & 22.5                                                                                        & 90.788                                                                                                                    & 2.3921                                                                                       & 500.73                                                                                            \\
CXOU J164710.2-455216 & 6.59                                                                                        & 26.591                                                                                                                    & 2.3921                                                                                       & 42.954                                                                                            \\
1RXS J170849.0-400910 & 46.8                                                                                        & 188.84                                                                                                                    & 2.3921                                                                                       & 2166.3                                                                                            \\
CXOU J171405.7-381031 & 50.1                                                                                        & 202.15                                                                                                                    & 2.3921                                                                                       & 2482.6                                                                                            \\
SGR J1745-2900        & 23.1                                                                                        & 93.209                                                                                                                    & 2.3921                                                                                       & 527.79                                                                                            \\
SGR 1806-20           & 196                                                                                         & 790.86                                                                                                                    & 2.3921                                                                                       & 37997                                                                                             \\
XTE J1810-197         & 21.0                                                                                        & 84.735                                                                                                                    & 2.3921                                                                                       & 436.19                                                                                            \\
Swift J1818.0-1607    & 35.4                                                                                        & 142.84                                                                                                                    & 2.3921                                                                                       & 1239.5                                                                                            \\
Swift J1822.3-1606    & 1.36                                                                                        & 5.4876                                                                                                                    & 2.3921                                                                                       & 1.8294                                                                                            \\
SGR 1833-0832         & 16.5                                                                                        & 66.578                                                                                                                    & 2.3921                                                                                       & 269.28                                                                                            \\
Swift J1834.9-0846    & 14.2                                                                                        & 57.297                                                                                                                    & 2.3921                                                                                       & 199.44                                                                                            \\
1E 1841-045           & 70.3                                                                                        & 283.66                                                                                                                    & 2.3921                                                                                       & 4888.2                                                                                            \\
3XMM J185246.6+003317 & 4.07                                                                                        & 16.422                                                                                                                    & 2.3921                                                                                       & 16.384                                                                                            \\
SGR 1900+14           & 70.0                                                                                        & 282.45                                                                                                                    & 2.3921                                                                                       & 4846.5                                                                                            \\
SGR 1935+2154         & 21.8                                                                                        & 87.963                                                                                                                    & 2.3921                                                                                       & 470.05                                                                                            \\
1E 2259+586           & 5.88                                                                                        & 23.726                                                                                                                    & 2.3921                                                                                       & 34.197                                                                                            \\
PSR J1846-0258 \#\#   & 4.88                                                                                        & 19.691                                                                                                                    & 2.3921                                                                                       & 23.555                                                                                            \\ \bottomrule
\end{tabular}
\label{tab-def}
\end{table}

\noindent 
In addition to studying various magnetar candidates and PSR B1257+12, we aim to estimate the contribution of the magnetic dipole to the classical tests for the Sun. It is known that the magnetic field at the poles of the solar photosphere, during periods of low activity, is approximately $10  \text{ }{\rm G}$ \cite{Duncan}. Accordingly, we calculate $\mu_{\odot}$ for $(B_{s} = 10^{-3}  \text{ }{\rm T}, R = 6.9634 \times 10^8  \text{ }{\rm m}, \theta = \pi)$. The Sun's angular momentum, $J = 1.8838 \times 10^{41} \text{ } {\rm kg} \text{ } {\rm m}^2 \text{ } {\rm s}^{-1}$, is taken from the calculations by Rongquin et al. \cite{Rongquin}. For both NSs and the Sun, we assume $Q_e = 0 \hspace{0.1cm} {\rm C}$.
Once all values are obtained, they are presented in tables that distinguish the contributions of each physical quantity and order of magnitude. This organization allows readers to evaluate the relative significance of the magnetic dipole’s contribution to the classical tests of General Relativity compared to those of mass, first-order rotation, etc.

\begin{table}[h]
\centering
\caption{Numerical estimates for the periastron precession for different magnetars and magnetar candidates of the McGill Magnetar Catalog, taking $M = 1.4 \, M_{\odot} , \, J/M = 1 \, {\rm m}, \, Q = -9648\, {\rm m^3}, \, Q_e = 0 \, {\rm m} $ and $r_{min}= r_0 = R=10 \, {\rm km}$. \cite{OK}}%
\begin{tabular}{@{}lllll@{}}
\toprule
Magnetar              & \multicolumn{1}{c}{\begin{tabular}[c]{@{}c@{}}Magnetic field \\ $10^9$  \rm T\end{tabular}} & \multicolumn{1}{c}{\begin{tabular}[c]{@{}c@{}}$\mu$ \\ $10^1$ $\rm m^2$ \end{tabular}} & \multicolumn{1}{c}{\begin{tabular}[c]{@{}c@{}}$ \Delta \varphi $\\ $10^1$ as/cent\end{tabular}} & \multicolumn{1}{c}{\begin{tabular}[c]{@{}c@{}}$\Delta \varphi_\mu$ \\ $10^{-30}$ as/cent \end{tabular}} \\ \midrule
CXOU J010043.1-721134 & 39.3                                                                                        & 158.58                                                                                                                    & 6.0129                                                                                       & 678.66                                                                                            \\
4U 0142+61            & 13.4                                                                                        & 54.069                                                                                                                    & 6.0129                                                                                       & 78.900                                                                                            \\
SGR 0418+5729         & 0.61                                                                                        & 2.4613                                                                                                                    & 6.0129                                                                                       & 0.1635                                                                                            \\
SGR 0501+4516         & 18.7                                                                                        & 75.455                                                                                                                    & 6.0129                                                                                       & 153.66                                                                                            \\
SGR 0526-66           & 56.0                                                                                        & 225.96                                                                                                                    & 6.0129                                                                                       & 1378.0                                                                                            \\
1E 1048.1-5937        & 38.6                                                                                        & 155.75                                                                                                                    & 6.0129                                                                                       & 654.70                                                                                            \\
1E 1547.0-5408        & 31.8                                                                                        & 128.31                                                                                                                    & 6.0129                                                                                       & 444.34                                                                                            \\
PSR J1622-4950        & 27.4                                                                                        & 110.56                                                                                                                    & 6.0129                                                                                       & 329.89                                                                                            \\
SGR 1627-41           & 22.5                                                                                        & 90.788                                                                                                                    & 6.0129                                                                                       & 222.45                                                                                            \\
CXOU J164710.2-455216 & 6.59                                                                                        & 26.591                                                                                                                    & 6.0129                                                                                       & 19.082                                                                                            \\
1RXS J170849.0-400910 & 46.8                                                                                        & 188.84                                                                                                                    & 6.0129                                                                                       & 962.40                                                                                           \\
CXOU J171405.7-381031 & 50.1                                                                                        & 202.15                                                                                                                    & 6.0129                                                                                       &  1102.9                                                                                           \\
SGR J1745-2900        & 23.1                                                                                        & 93.209                                                                                                                    & 6.0129                                                                                       & 234.47                                                                                            \\
SGR 1806-20           & 196                                                                                         & 790.86                                                                                                                    & 6.0129                                                                                       & 16880                                                                                             \\
XTE J1810-197         & 21.0                                                                                        & 84.735                                                                                                                    & 6.0129                                                                                       & 193.78                                                                                            \\
Swift J1818.0-1607    & 35.4                                                                                        & 142.84                                                                                                                    & 6.0129                                                                                       & 550.64                                                                                            \\
Swift J1822.3-1606    & 1.36                                                                                        & 5.4876                                                                                                                    & 6.0129                                                                                       & 0.8127                                                                                            \\
SGR 1833-0832         & 16.5                                                                                        & 66.578                                                                                                                    & 6.0129                                                                                       & 119.63                                                                                            \\
Swift J1834.9-0846    & 14.2                                                                                        & 57.297                                                                                                                    & 6.0129                                                                                       & 88.602                                                                                           \\
1E 1841-045           & 70.3                                                                                        & 283.66                                                                                                                    & 6.0129                                                                                       & 2171.6                                                                                            \\
3XMM J185246.6+003317 & 4.07                                                                                        & 16.422                                                                                                                    & 6.0129                                                                                       & 7.2787                                                                                            \\
SGR 1900+14           & 70.0                                                                                        & 282.45                                                                                                                    & 6.0129                                                                                       & 2153.1                                                                                            \\
SGR 1935+2154         & 21.8                                                                                        & 87.963                                                                                                                    & 6.0129                                                                                       & 208.82                                                                                            \\
1E 2259+586           & 5.88                                                                                        & 23.726                                                                                                                    & 6.0129                                                                                       & 15.192                                                                                            \\
PSR J1846-0258 \#\#   & 4.88                                                                                        & 19.691                                                                                                                    & 6.0129                                                                                       & 10.464                                                                                            \\ \bottomrule
\end{tabular}
\label{tab-pres}
\end{table}

\noindent 
Table \ref{inv} shows the contributions of quantities independent of the magnetic dipole to the classical tests. Since we assume identical masses, radii, and angular momenta for all magnetars, these quantities are the same across all magnetars, which is why each contribution $(\Delta \phi , \, \Delta \varphi , \, \Delta t , \, z)$ is represented by a single value. From now on, when discussing the significance of the magnetic dipole moment in classical tests, these values from Table \ref{inv} serve as the point of comparison.

\noindent 
As expected, the mass contribution to the various tests is much greater than the contributions from other physical quantities. However, for the deflection angle, both the second-order rotation and first-order quadrupole moment contributions are on the order of milliarcseconds (mas). This is noteworthy since the angular resolution of the GAIA EDR3 is about $0.5 \, {\rm mas}$ for objects of magnitude $20$, which is comparable to the objects studied here \cite{GAIA, OK}. For the other classical tests, the contributions from $J , \, Q$, or combinations of these quantities are negligible compared to the total value.

\noindent 
The use of specific values here aims to provide a clearer understanding of the orders of magnitude of each physical quantity's contribution to the total value of the classical tests. For all tests, we ensured that the relationships between the magnitudes of different terms remained consistent for varying values of $(r, \, r_0)$.

\begin{table}[h]
\centering
\caption{Numerical estimates for the time delay for different magnetars and magnetar candidates of the McGill Magnetar Catalog, taking $M = 1.4 \, M_{\odot} , \, J/M = 1 \, {\rm m}, \, Q = -9648\, {\rm m^3}, \, Q_e = 0 \, {\rm m} $ and $r_{min}= r_0 = R=10 \, {\rm km}$. \cite{OK}}%
\begin{tabular}{@{}lllll@{}}
\toprule
Magnetar              & \multicolumn{1}{c}{\begin{tabular}[c]{@{}c@{}}Magnetic field \\ $10^9$  \rm T\end{tabular}} & \multicolumn{1}{c}{\begin{tabular}[c]{@{}c@{}}$\mu$ \\ $10^1$ $\rm m^2$ \end{tabular}} & \multicolumn{1}{c}{\begin{tabular}[c]{@{}c@{}}$ \Delta t $\\ $10^{-4}$ s\end{tabular}} & \multicolumn{1}{c}{\begin{tabular}[c]{@{}c@{}}$\Delta t_\mu$ \\ $10^{-18}$ s \end{tabular}} \\ \midrule
CXOU J010043.1-721134 & 39.3                                                                                        & 158.58                                                                                                                    & 9.6503                                                                                       & 823.48                                                                                            \\
4U 0142+61            & 13.4                                                                                        & 54.069                                                                                                                    & 9.6503                                                                                       & 95.736                                                                                            \\
SGR 0418+5729         & 0.61                                                                                        & 2.4613                                                                                                                    & 9.6503                                                                                       & 0.1984                                                                                            \\
SGR 0501+4516         & 18.7                                                                                        & 75.455                                                                                                                    & 9.6503                                                                                       & 186.44                                                                                            \\
SGR 0526-66           & 56.0                                                                                        & 225.96                                                                                                                    & 9.6503                                                                                       & 1672.0                                                                                            \\
1E 1048.1-5937        & 38.6                                                                                        & 155.75                                                                                                                    & 9.6503                                                                                       & 794.41                                                                                            \\
1E 1547.0-5408        & 31.8                                                                                        & 128.31                                                                                                                    & 9.6503                                                                                       & 539.17                                                                                            \\
PSR J1622-4950        & 27.4                                                                                        & 110.56                                                                                                                    & 9.6503                                                                                       & 400.29                                                                                            \\
SGR 1627-41           & 22.5                                                                                        & 90.788                                                                                                                    & 9.6503                                                                                       & 269.92                                                                                            \\
CXOU J164710.2-455216 & 6.59                                                                                        & 26.591                                                                                                                    & 9.6503                                                                                       & 23.155                                                                                            \\
1RXS J170849.0-400910 & 46.8                                                                                        & 188.84                                                                                                                    & 9.6503                                                                                       & 1167.8                                                                                           \\
CXOU J171405.7-381031 & 50.1                                                                                        & 202.15                                                                                                                    & 9.6503                                                                                       &  1338.3                                                                                           \\
SGR J1745-2900        & 23.1                                                                                        & 93.209                                                                                                                    & 9.6503                                                                                       & 284.51                                                                                            \\
SGR 1806-20           & 196                                                                                         & 790.86                                                                                                                    & 9.6503                                                                                       & 20482                                                                                             \\
XTE J1810-197         & 21.0                                                                                        & 84.735                                                                                                                    & 9.6503                                                                                       & 235.13                                                                                            \\
Swift J1818.0-1607    & 35.4                                                                                        & 142.84                                                                                                                    & 9.6503                                                                                       & 668.15                                                                                            \\
Swift J1822.3-1606    & 1.36                                                                                        & 5.4876                                                                                                                    & 9.6503                                                                                       & 0.9862                                                                                            \\
SGR 1833-0832         & 16.5                                                                                        & 66.578                                                                                                                    & 9.6503                                                                                       & 145.16                                                                                            \\
Swift J1834.9-0846    & 14.2                                                                                        & 57.297                                                                                                                    & 9.6503                                                                                       & 107.51                                                                                           \\
1E 1841-045           & 70.3                                                                                        & 283.66                                                                                                                    & 9.6503                                                                                       & 2635.0                                                                                            \\
3XMM J185246.6+003317 & 4.07                                                                                        & 16.422                                                                                                                    & 9.6503                                                                                       & 8.8320                                                                                            \\
SGR 1900+14           & 70.0                                                                                        & 282.45                                                                                                                    & 9.6503                                                                                       & 2612.5                                                                                            \\
SGR 1935+2154         & 21.8                                                                                        & 87.963                                                                                                                    & 9.6503                                                                                       & 253.38                                                                                            \\
1E 2259+586           & 5.88                                                                                        & 23.726                                                                                                                    & 9.6503                                                                                       & 18.434                                                                                            \\
PSR J1846-0258 \#\#   & 4.88                                                                                        & 19.691                                                                                                                    & 9.6503                                                                                       & 12.697                                                                                            \\ \bottomrule
\end{tabular}
\label{tab-time}
\end{table}

\noindent 
Table \ref{tab-def} presents the deflection angle of light for various magnetars from the McGill Magnetar Catalog \cite{OK}. The first column lists the system's name, followed by the magnetic field, the corresponding value of $\mu$, the total deflection angle, and the first- and second-order contributions of the magnetic dipole to the deflection angle. An interesting result is that $\Delta \phi_{\mu}$ is, on average, two orders of magnitude smaller than both $\Delta \phi_{J2}$ and $\Delta \phi_{Q1}$, meaning that even for a magnetar, characterized by slow rotation and a higher magnetic field than normal NS, the contribution from the quadrupole moment $\Delta \phi_{Q1}$ is still more significant. For the strong magnetic field of SGR 1806-20, $\Delta \phi_{\mu}$ is still one order of magnitude smaller than $\Delta \phi_{J2}$ and $\Delta \phi_{Q1}$.

\noindent 
Although the precise angular momentum of these systems is uncertain, we performed the calculations using an upper bound for $J$, ensuring the rotation remains slow. The relationship between the magnetic field and the square root of the period suggests that magnetars with extremely strong magnetic fields likely have slower rotations.

\noindent 
Our results for the magnetic dipole contribution to light deflection differ from those obtained by Kim, who found a result on the order of mas \cite{Kim}. While the approaches and metrics differ, we attribute the discrepancy to different methodologies in calculating the magnetic dipole moment.

\noindent 
The results for the periastron precession and time delay are presented in Tables \ref{tab-pres} and \ref{tab-time}, respectively. The structure of these tables follows the same format described for Table \ref{tab-def}. The values in Table \ref{tab-pres} were derived by estimating the periastron precession for a Mercury-like object orbiting a magnetar. For this scenario, the semi-latus rectum is taken as $l=a(1-e^2) \sim 0.37 \, {\rm AU}$, and the specific angular momentum as $l_z = a(1-e)v_{p} \sim 2.74 \times 10^{15} \, {\rm m}^2 \, {\rm s}^{-1}$, where $v_{p}$ is Mercury's periastron speed. Although these values may seem arbitrary, care was taken to preserve realistic relationships between their orders of magnitude. Consequently, any other plausible choices for the semi-latus rectum and specific angular momentum would yield consistent conclusions. It is also worth noting that PSR B1257+12 A is a Mercury-like exoplanet orbiting a millisecond pulsar \cite{Wolszczan}.

\begin{table}[h]
\centering
\caption{Numerical estimates for the mass, angular momentum, magnetic dipole moment and quadrupole moment contributions to the angle of light deflection, periastron precession and time delay for the PSR B1257+12, taking $\mu = 34.52 \, {\rm m^2}$, $M = 1.4 M_{\odot}$, $J/M = 255{\rm m} $, $Q = -488419275.09 \,{\rm m^3}$, $Q_e = 0 \, {\rm m}$  and $r_{min}=r_0=R=10 \, {\rm km}$.}%
\begin{tabular}{@{}lllll@{}}
\toprule
& M $1^{\text{st}}$ order & M $2^{\text{nd}}$ order & J $1^{\text{st}}$ order   \\
\midrule
$\Delta \phi$ (as) 
& $1.70613
 \times10^{5}$       & $6.8630\times10^{4}$        &  $3.8388\times10^{3}$     \\
$\Delta \varphi$ (as/cent) & $6.1942\times10^2$        & $1.0706\times10^{-5}$     & \hspace{0.5cm} -    \\
$\Delta t$ (s)             &  $9.4459\times 10^{-4}$    & $1.2603\times 10^{-5}$    &   $6.2080\times 10^{-7}$     \\
z  &2.0679$\times10^{-1}$        & $6.4142\times10^{-2}$     & \hspace{0.5cm} -    \\
\bottomrule
\end{tabular}

\begin{tabular}{@{}lllll@{}}
\toprule
& J $2^{\text{nd}}$ order & Q $1^{\text{st}}$ order &  Q $2^{\text{nd}}$ order  \\
\midrule
$\Delta \phi$ (as) 
& $1.2120\times10^{2}$       &   $4.0297\times10^{2}$        & $2.6095\times10^{-1}$    \\
$\Delta \varphi$ (as/cent) & $8.2340\times10^{-13}$        &  2.5004$\times 10^{-5}$     & $1.7445\times10^{-18}$    \\
$\Delta t$ (s)             &  $7.0892\times 10^{-11}$    &   $3.2584\times 10^{-8}$    &  $4.1013\times 10^{-12}$     \\
z  &  \hspace{0.5cm} -        & $2.4421\times10^{-4}$     & $2.9819\times10^{-8}$    \\
\bottomrule
\end{tabular}

\begin{tabular}{@{}lllll@{}}
\toprule
& JM $1^{\text{st}}$ order & JQ $1^{\text{st}}$ order &  MQ $1^{\text{st}}$ order  \\
\midrule
$\Delta \phi$ (as) 
&  $3.0594\times10^{3}$       & $1.1276\times10^{1}$   & $4.3364\times10^{2}$     \\
$\Delta \varphi$ (as/cent) &  $ 1.3073\times10^{1}$        & $1.7207\times 10^{-4}$     &  $1.4406\times 10^{-11}$    \\
$\Delta t$ (s)             &   $2.0165\times 10^{-7}$    &  $4.4651\times 10^{-11}$    &  $2.1499\times 10^{-8}$     \\
z  &\hspace{0.5cm} -        & \hspace{0.5cm} -     & $2.0199\times10^{-4}$   \\
\bottomrule
\end{tabular}

\begin{tabular}{@{}llll@{}}
\toprule
& $\mu$ $2^{\text{nd}}$ order & Total  \\
\midrule
$\Delta \phi$ (as) 
&  $7.2385\times10^{-9}$       & $2.3329\times10^{5}$      \\
$\Delta \varphi$ (as/cent) &  $3.6216\times10^{-31}$        & $7.5016\times 10^{2}$         \\
$\Delta t$ (s)             &   $3.9020\times 10^{-19}$    &  $9.5643\times 10^{-4}$        \\
z  &\hspace{0.5cm} -        & $2.7049\times 10^{-1}$     \\
\bottomrule
\end{tabular}
\label{tab-PSR}
\end{table}

\noindent 
Notably, even in this extreme magnetar scenario, $\Delta \varphi _{\mu}$ is, at best, $14$ orders of magnitude smaller than $\Delta \varphi _{Q1}$ and 6 orders of magnitude smaller than $\Delta \varphi _{J2}$ (as there is no first-order contribution from $J$). This demonstrates that periastron precession is significantly more sensitive to rotational effects than to the magnetic dipole moment, especially when compared to the light deflection angle.

\noindent 
A comparable trend is observed in the time delay results presented in Table \ref{tab-time}. These values were estimated for a light emitter at a distance similar to Venus' orbit and a receiver at a distance similar to Earth's orbit around the magnetar, as illustrated in Figure \ref{time-delay}. Under these conditions, the distances $D_{1}$ and $D_{2}$ were set to $0.72 \, {\rm AU}$ and  $1 \, {\rm AU} $, respectively. Our tests confirm that varying $D_1$ or $D_2$ does not affect the qualitative results. In this case, $\Delta t_{\mu}$ lies between $\Delta t_{J 2}$ and $\Delta t_{Q 1}$, which contrasts with the dominance patterns observed in periastron precession and highlights an interesting aspect of time delay effects.

\noindent
Table \ref{tab-PSR} shows the results of the classical tests applied to PSR B1257+12. The periastron precession was calculated with the exoplanet PSR B1257+12 B, which means that the keplerian parameters where taken from Konacki and Wolszczan \cite{Konacki}. The specific angular momentum was calculated to be $l_z \sim 3.16\times 10^{15} \text{ m}^2 \text{s}^{-1}$. The results for the contributions on the time delay are between  a hypothetical light emitter at PSR B1257+12 B and a receiver at PSR B1257+12 C, which means that $D_1$ and $D_2$ were set to $0.36$ AU and $0.46$ AU respectively. PSR B1257+12 A was not considered, even though it resembles more a Mercury-like planet, because it lacks a measurable eccentricity. The results for the contributions of rotation and quadrupole moment are expected if one takes into account that angular momentum is around 2 orders of magnitude greater than in the magnetars case, which also means that the quadrupole moment is 4 orders of magnitude greater. This should impact in 2 orders of magnitude the results for $J$ in the first order and in 4 orders of magnitude the results for $Q$ in the first order, which is what we found. As a result of this, there is a stronger argument for detection of the rotation and quadrupole moment contributions on the light deflection in millisecond pulsars than magnetars. This may seem obvious at first, because millisecond pulsars have higher angular momentum. However, we also found that the contributions of rotation and quadrupole moment are significant enough to the total to be plausibly measurable. For the other classical tests, the rotational and quadrupole moment contributions are less significant to the total, with the exception of the $JM$ first order of the periastron precession. The magnetic dipole effect in all classical tests is also negligible in this case.  

\noindent 
Table \ref{tab-Sun} presents the corresponding results for the classical tests applied to the Sun. It is crucial to note that the magnetic dipole approximation for the Sun's photospheric magnetic field is accurate only during periods of low activity. The previously mentioned value for the Sun's angular momentum, $J_{\odot}$, corresponds to a spin parameter $J/M \sim 315 \, {\rm m}$. Using this, the massive quadrupole moment was estimated via equation Eq.(\ref{QJ}). If this $J/M$
value appears counterintuitive, recall that the Sun's mass is distributed over $6.96\times 10^{5} \, {\rm km}$, rather than the $10 \, {\rm km}$ typical of compact objects, and angular momentum depends on the square of the distance between the axis of rotation and the mass.

\noindent 
For light deflection, the contribution from the magnetic dipole is significantly smaller than that from rotation, and both are dwarfed by the contribution from mass. In all classical tests applied to the Sun during periods of low activity, the effects of the magnetic dipole moment are found to be negligible and beyond measurable limits.

\begin{table}[h]
\centering
\caption{Numerical estimates for the mass, angular momentum, magnetic dipole moment and quadrupole moment contributions to the angle of light deflection, periastron precession and time delay for the Sun in epochs of low activity, taking $\mu = 13652.45 \, {\rm m^2}$, $M = M_{\odot}$, $J = J_{\odot}$, $Q = -689190574.23 \,{\rm m^3}$  and $r_{min}=r_0=R=R_{\odot}$.}%
\begin{tabular}{@{}lllll@{}}
\toprule
& M $1^{\text{st}}$ order & M $2^{\text{nd}}$ order & J $1^{\text{st}}$ order   \\
\midrule
$\Delta \phi$ (as) 
& $1.7502 \times10^{0}$       & $7.2221\times10^{-6}$        &  $7.9409\times10^{-7}$     \\
$\Delta \varphi$ (as/cent) & $4.2949\times10^1$        & $5.1473\times10^{-6}$     & \hspace{0.5cm} -    \\
$\Delta t$ (s)             &  $9.0907\times 10^{-4}$    & $9.2210\times 10^{-11}$    &   $8.9416\times 10^{-12}$     \\
z  &2.1213$\times10^{-6}$        & $6.7499\times10^{-12}$     & \hspace{0.5cm} -    \\
\bottomrule
\end{tabular}

\begin{tabular}{@{}lllll@{}}
\toprule
& J $2^{\text{nd}}$ order & Q $1^{\text{st}}$ order &  Q $2^{\text{nd}}$ order  \\
\midrule
$\Delta \phi$ (as) 
& $5.1865\times10^{-18}$       &   $1.6844\times10^{-12}$        & $4.5590\times10^{-30}$    \\
$\Delta \varphi$ (as/cent) & $2.7333\times10^{-15}$        &  1.1995$\times 10^{-7}$     & $4.0150\times10^{-23}$    \\
$\Delta t$ (s)             &  $2.1197\times 10^{-25}$    &   $9.4831\times 10^{-18}$    &  $5.0068\times 10^{-36}$     \\
z  &  \hspace{0.5cm} -        & $1.0208\times10^{-18}$     & $5.2097\times10^{-37}$    \\
\bottomrule
\end{tabular}

\begin{tabular}{@{}lllll@{}}
\toprule
& JM $1^{\text{st}}$ order & JQ $1^{\text{st}}$ order &  MQ $1^{\text{st}}$ order  \\
\midrule
$\Delta \phi$ (as) 
&  $6.4921\times10^{-12}$       & $9.7493\times10^{-24}$   & $1.8594\times10^{-17}$     \\
$\Delta \varphi$ (as/cent) &  $1.9412\times10^{-3}$        & $4.8994\times 10^{-8}$     &  $4.7919\times 10^{-14}$    \\
$\Delta t$ (s)             &   $2.9900\times 10^{-17}$    &  $2.6977\times 10^{-30}$    &  $6.4412\times 10^{-23}$     \\
z  &\hspace{0.5cm} -        & \hspace{0.5cm} -     & $8.6613\times10^{-24}$   \\
\bottomrule
\end{tabular}

\begin{tabular}{@{}llll@{}}
\toprule
& $\mu$ $2^{\text{nd}}$ order & Total  \\
\midrule
$\Delta \phi$ (as) 
&  $4.8171\times10^{-23}$       & $1.7502\times10^{0}$      \\
$\Delta \varphi$ (as/cent) &  $5.0303\times10^{-26}$        & $4.2951\times 10^{1}$         \\
$\Delta t$ (s)             &   $1.8144\times 10^{-28}$    &  $9.0907\times 10^{-4}$        \\
z  &\hspace{0.5cm} -        & $2.1213\times 10^{-6}$     \\
\bottomrule
\end{tabular}
\label{tab-Sun}
\end{table}

\section{Conclusions}

In previous years, there has been an interest in studying in more detail the effect the magnetic dipole moment can have in the deflection of light for a system like a magnetar and in quantifying this effect. By employing the HT-like metric with magnetic dipole moment, massive quadrupole moment and electric charge, we have derived expressions that account for the mass, electromagnetic, and rotational effects. This has allowed us to obtain an estimation on the orders of magnitude of the contribution of all classical tests. Then, the McGill Magnetar Catalog is used to estimate numerically the contribution of the magnetic dipole to the angle of light deflection, the periastron precession, and the time delay, and we have observed that the magnetic dipole effect can be quite comparable, and even greater than the rotational contribution depending on the analyzed test and physical conditions.

\noindent 
For magnetars, the results for light deflection indicate that the magnetic dipole contribution is approximately two orders of magnitude smaller than the rotational contribution. In the case of periastron precession and time delay, the magnetic dipole moment contributes but is about six orders of magnitude smaller than the second-order rotational contribution for periastron precession and five orders of magnitude smaller for time delay. Regarding gravitational redshift, the magnetic dipole contribution is found to be negligible within the approximation used in this study.

\noindent 
The analysis was also extended to a specific millisecond pulsar (PSR B1257+12). While the rotational and quadrupole moment contributions are more significant, the magnetic dipole contribution remains negligible. However, it is possible that the approximation presented may not be entirely applicable to millisecond pulsars due to their higher angular momentum. Additionally, while the analysis of Laarakkers and Poisson is employed to provide a reasonable estimation of the quadrupole moments for the systems studied, we acknowledge that alternative relationships can estimate the quadrupole moment as a function of angular momentum and magnetic stress, potentially amplifying its contribution. Exploring theoretical limits, such as the maximum deformation constrained by the Virial limit \cite{Suvorov}, could further amplify the effects of the massive quadrupole moment.

\noindent 
A more detailed investigation into the possible range of masses, radii, angular momenta, and their relationship to the massive quadrupole moment—accounting for the EoS of neutron stars—is critical for advancing this research. Another interesting and important avenue to explore is the effects in binary systems of compact objects, which require a different formalism (we encourage interested readers to consult \cite{Handbook, Damour, Bagchi}).

\noindent 
Finally, values were calculated for the Sun during periods of low activity. For these conditions, none of the classical tests indicated a significant contribution from the magnetic dipole moment.

\bigskip
\noindent
{\textbf{Aknowledgement}:}

\noindent 
We want to thank Dr. Heidy Guti\'errez-Garro for helping us with information regarding the solar magnetic field.




\appendixpage

\appendix

\section{Solution of the differential equation}

The other 23 equations are given by

\begin{eqnarray}\label{difeqs}
-\frac{d^2 u_{e21}}{d\varphi^2} - u_{e21} - 2 \cos^3(\varphi) &=& 0, \nonumber\\
-\frac{d^2 u_{e\mu 4}}{d\varphi^2} - u_{e\mu 4} + 2 \cos(\varphi) &=& 0, \nonumber\\
-\frac{d^2 u_{J14}}{d\varphi^2} - u_{J14} - 2 &=& 0, \nonumber\\
-\frac{d^2 u_{J21}}{d\varphi^2} - u_{J21} - \frac{81}{2} \cos^5(\varphi) &=& 0, \nonumber\\
-\frac{d^2 u_{J23}}{d\varphi^2} - u_{J23} + 34 \cos^3(\varphi) &=& 0, \nonumber\\
-\frac{d^2 u_{J25}}{d\varphi^2} - u_{J25} + 12 \cos(\varphi) &=& 0, \nonumber\\
-\frac{d^2 u_{JM4}}{d\varphi^2} - u_{JM4} + 6 \cos(\varphi) u_{J14} - 8 \cos(\varphi) &=& 0, \nonumber\\
-\frac{d^2 u_{JM5}}{d\varphi^2} - u_{JM5} + 6 &=& 0, \nonumber\\
-\frac{d^2 u_{JQ4}}{d\varphi^2} - u_{JQ4} - 18 \cos^3(\varphi) + 6 \cos(\varphi) u_{J12} &=& 0, \nonumber\\
-\frac{d^2 u_{JQ6}}{d\varphi^2} - u_{JQ6} + 6 \cos(\varphi) u_{J14} &=& 0, \nonumber\\
-\frac{d^2 u_{JQ7}}{d\varphi^2} - u_{JQ7} + 6 &=& 0, \\
-\frac{d^2 u_{M11}}{d\varphi^2} - u_{M11} + 3 \cos^2(\varphi) &=& 0, \nonumber\\
-\frac{d^2 u_{M21}}{d\varphi^2} - u_{M21} + 6 \cos(\varphi) u_{M11} &=& 0, \nonumber\\
-\frac{d^2 u_{\mu 21}}{d\varphi^2} - u_{\mu 21} - \frac{7}{2} \cos^5(\varphi) &=& 0, \nonumber\\
-\frac{d^2 u_{\mu 23}}{d\varphi^2} - u_{\mu 23} + \frac{8}{3} \cos^3(\varphi) &=& 0, \nonumber\\
-\frac{d^2 u_{Q13}}{d\varphi^2} - u_{Q13} + 3 \cos^2(\varphi) &=& 0, \nonumber\\
-\frac{d^2 u_{Q21}}{d\varphi^2} - u_{Q21} + 33 \cos^7(\varphi) &=& 0, \nonumber\\
-\frac{d^2 u_{Q23}}{d\varphi^2} - u_{Q23} - \frac{93}{4} \cos^5(\varphi) + 6 \cos(\varphi) u_{Q11} &=& 0, \nonumber\\
-\frac{d^2 u_{Q25}}{d\varphi^2} - u_{Q25} + 6 \cos(\varphi) u_{Q13} &=& 0, \nonumber\\
-\frac{d^2 u_{Q26}}{d\varphi^2} - u_{Q26} - 6 \cos^2(\varphi) + 6 \cos(\varphi) u_{Q14} &=& 0, \nonumber\\
-\frac{d^2 u_{QM1}}{d\varphi^2} - u_{QM1} + \frac{3}{2} \cos^5(\varphi) + 6 \cos(\varphi) u_{Q11} &=& 0, \nonumber\\
-\frac{d^2 u_{QM3}}{d\varphi^2} - u_{QM3} + 10 \cos^3(\varphi) + 6 \cos(\varphi) u_{M11} + 6 \cos(\varphi) u_{Q13} &=& 0, \nonumber\\
-\frac{d^2 u_{QM4}}{d\varphi^2} - u_{QM4} - 6 \cos^2(\varphi) + 6 \cos(\varphi) u_{M12} + 6 \cos(\varphi) u_{Q14} &=& 0.\nonumber
\end{eqnarray}

\noindent 
The solutions are

\begin{align}
\mathcal{R}_1  &=  u_{e21} = \frac{-1}{17} u_{J23} = \frac{1}{9} u_{JQ4} = \frac{-3}{4} u_{\mu 23}, \nonumber \\
\mathcal{R}_2  &= u_{e\mu 4} = \frac{1}{6}u_{J25}=\frac{-1}{10}u_{JM4}=\frac{-1}{6}u_{JQ6}, \nonumber \\
\mathcal{R}_3  &= u_{J14} = \frac{-1}{3}u_{JM5}=\frac{-1}{3} u_{JQ7}, \nonumber \\
\mathcal{R}_4  &=  u_{J21} = \frac{27}{7}u_{\mu 21}=\frac{54}{31}u_{Q23}=-27u_{QM1},  \\
\mathcal{R}_5  &= u_{M11} =u_{Q13}=\frac{-1}{2}u_{Q26}=\frac{-1}{2}u_{QM4}, \nonumber \\
\mathcal{R}_6  &= u_{M21} =u_{Q25}= \frac{1}{6}(2 u_{QM3}+\cos(3\varphi)), \nonumber \\
\mathcal{R}_7  &= u_{Q21}  . \nonumber
\end{align}

\section{Gravitational potential of HT-like metric}

\begin{align}\label{E2}
\frac{\varepsilon^2 - 1}{2} &= -\left(\frac{M}{l}\right)  + \frac{1}{2} \left(\frac{l_z}{l}\right)^2 + \frac{1}{2} \left(\frac{Q_e}{l}\right)^2  - l_z^2 \left(\frac{M}{l^3}\right)  \nonumber \\
& - \frac{1}{2} \left(\frac{Q}{l^3}\right)  - 2 l_z \left(\frac{J}{l^3}\right)  - \frac{1}{2} \left(\frac{Q M}{l^4}\right)  + 2 l_z \left(\frac{J M}{l^4}\right) \nonumber \\
&+ 2 \left(\frac{J^2}{l^4}\right)  + \frac{l_z^2}{2} \left(\frac{ Q_e^2}{l^4}\right) + l_z \left(\frac{\mu Q_e}{l^4}\right)  - l_z^2 \left(\frac{Q}{l^5}\right) \\
&- l_z^3 \left(\frac{J}{l^5}\right)   - \frac{3 l_z^2}{4} \left(\frac{QM}{l^6}\right) + \frac{1}{4} \left(\frac{Q^2}{l^6}\right)  + l_z^3 \left(\frac{J M}{l^6}\right)  \nonumber \\
&+ \frac{9 l_z}{2} \left(\frac{J Q}{l^6}\right)  + \frac{17 l_z^2}{4} \left(\frac{J^2}{l^6}\right)  + \frac{l_z^3}{2} \left(\frac{\mu Q_e}{l^6}\right)   - \frac{l_z^2}{12} \left(\frac{\mu^2}{l^6}\right) \nonumber\\
&+ {\cal O}(M^3, \, J^3,\, Q^3, \, \mu^3, \, Q_e^3) \nonumber.
\end{align}

\section{Gutsunaev-Manko metric}

In this appendix, we analyze the metric proposed by Gutsunaev and Manko \cite{GM}. This exact metric originally has a magnetic dipole and does not have rotation, and reduces to Schwarzschild if the magnetic dipole is turned off. This metric in prolate spheroidal coordinates is

\begin{eqnarray}
d s^2 = - f dt^2 + \frac{k^2}{f} {(x^2 - y^2) {\rm e}^{2 \gamma}} \left(\frac{d x^2}{x^2 - 1} + \frac{d y^2}{1 - y^2} \right) + \frac{k^2}{f} (x^2 - 1)(1 - y^2) d \phi^2 , 
\end{eqnarray}

\noindent
where

\begin{eqnarray}
f &=& \left(\frac{x - 1}{x + 1}\right) \left[\frac{z^2 + 4 \alpha^2 x^2 (1 - y^2)}{z^2 - 4 \alpha^2 y^2 (x^2 - 1)} \right]^2 \nonumber, \\
{\rm e}^{2 \gamma} &=& \left(\frac{x^2 - 1}{x^2 - y^2}\right)
\left[\frac{(z^2 + 4 \alpha^2 x^2 (1 - y^2))^4}{[(1 + \alpha^2) (x^2 - y^2)]^8} \right]^2 ,
\end{eqnarray}

\noindent
and $k$ and $\alpha$ are parameters related with the mass $m$ and the magnetic dipole $\mu$ of the object through 

\begin{eqnarray}
k &=& m \frac{(1 + \alpha^2)}{(1 - 3 \alpha^2)}, \nonumber \\ 
\mu &=& \frac{8 m^2 \alpha^3}{(1 - 3 \alpha^2)^2}, \\
z&=& x^2 - y^2 + \alpha^2 (x^2 - 1). \nonumber
\end{eqnarray}

\noindent 
The only component of the four potential $ A_{\mu} $ is 

\begin{eqnarray}
\label{Ap}
A_{\varphi} &=& \frac{4 k \alpha^3}{(1 + \alpha^2)} 
\left[\frac{(1 - y^2)\hat{z}}{z^2 + 4 \alpha^2 x^2 (1 - y^2)} \right] , 
\end{eqnarray}

\noindent 
where

\begin{eqnarray}
\hat{z} &=& 2(1 + \alpha^2) x^3 + (1 - 3 \alpha^2) x^2 + y^2 + \alpha^2. \nonumber 
\end{eqnarray}

\subsection{Spherical coordinates}

\noindent 
The GM metric components in spherical coordinates $(r, \, \theta, \, \phi)$ are obtained by means of the following transformation

\begin{eqnarray}
k x &=& r - m , \nonumber \\
y &=& \cos{\theta} . \\
\end{eqnarray}

\noindent 
The metric takes the form

\begin{equation}
\label{metric}
ds^2 = -V dt^2  + 2W dt d\varphi + X dr^2 + Y d \theta^2 + Z d \varphi^2 ,
\end{equation}

\noindent 
where the metric potentials are

\begin{align}
V &= f , \nonumber \\
W &= 0 , \nonumber \\
X &= \frac{\beta {\rm e}^{2 \gamma}}{\eta f} , \nonumber \\
Y &= \eta X , \\
Z &= \frac{\eta}{f} \sin^2{\theta} , \nonumber \\
A_{\varphi} &= \frac{4 m^2 \alpha^3}{(1-3 \alpha^2)} \frac{P}{M} \sin^2{\theta} , \nonumber 
\end{align}

\noindent
with

\begin{align}
\eta &= (r - m)^2-\frac{m^2 \left(1 + \alpha^2 \right)^2}{\left(1 - 3 \alpha^2 \right)^2} , \nonumber \\
\beta &= (r - m)^2-\frac{m^2 \left(1 + \alpha^2 \right)^2}{\left(1 - 3 \alpha^2 \right)^2} \cos^2{\theta} , \nonumber  \\
\lambda &= \Bigg((1-3\alpha^2)(r-m)- m \alpha^2 \Bigg)^2, \nonumber \\
\xi &=  m^2 (\alpha^2 - (1 + \alpha^2) \cos^2{\theta}), \nonumber\\
\sigma &= \left(1 - 3 \alpha^2\right)^3 (2 r - m)(r - m)^2\nonumber\\
\varepsilon&= 4 \alpha^2 m^2 (1 - 3 \alpha^2)^{2} (r - m)^2 \sin^2{\theta}, \\
{\rm e}^{2 \gamma} &= \frac{(1 - 3 \alpha^2)^2 \eta M^4}{((1 - 3 \alpha^2)^2 \beta)^9} , \nonumber \\
f &= \Bigg(1 - \frac{2 m \left(1 + \alpha^2 \right)}{\left(1 - 3 \alpha^2 \right) r + 4 m \alpha^2} \Bigg) \frac{M^2}{N^2} , \nonumber \\
P &= \sigma + m^3 \left(1 + \alpha^2\right)^2 \left(\alpha^2 + \cos^2{\theta} \right), \nonumber \\
M &= \left[(1 - 3 \alpha^2)^{2} \beta -\alpha m^2 (1 + \alpha^2) \right]^2 + \varepsilon, \nonumber \\
N &= (\lambda+\xi)^2 - 4 m^2 \alpha^2 (1 - 3 \alpha^2)^2 \eta \cos^2{\theta} . \nonumber 
\end{align}

\noindent
When testing these expressions to make sure they satisfy the EME, we found that there was a misprint in the function $f$ in spherical coordinates in the original article \cite{GM}.

\subsection{\label{sec:citeref} Taylor Series}

Let us expand the metric potentials in Taylor series up to order 
$ {\cal O}({\cal U}^3, \, J^2, \, \alpha^5) $:

\begin{align}
\label{expanmet}
V &= 1 - 2 {\cal U} , \nonumber \\
X &= 1 + 2 {\cal U} + 4 (1 + 2 \alpha^2 (1 + 5 \alpha^2) \sin^2{\theta}) {\cal U}^2 , \\
Y &= r^2 (1 - 8 (1 + 5 \alpha^2) \alpha^2 {\cal U}^2 \cos^2{\theta}) , \nonumber \\
Z &= r^2 (1 - 8 (1 + 5 \alpha^2) \alpha^2 {\cal U}^2) \sin^2{\theta} , \nonumber \\
A_{\varphi} &= 8 \alpha^3 m {\cal U} \sin^2{\theta} = (1 - 3 \alpha^2)^2 \mu {\cal U} \sin^2{\theta} , \nonumber 
\end{align}

\noindent
where $ {\cal U} = m/r $, and $J$ is the angular momentum. The magnetic dipole moment is given by

\begin{equation}
\label{mualpha}
\mu = 8m^2\alpha^3 + {\cal O}(m^6, \, \alpha^5) . 
\end{equation}

\noindent
From Eq.(\ref{mualpha})

\begin{equation}
\alpha \simeq \left(\frac{\mu}{8 m^2} \right)^{1/3}
\end{equation}

\noindent
The terms in the metric potentials Eq.(\ref{expanmet}) have square powers

$$ \alpha^2 = \left(\frac{\mu}{8 m^2} \right)^{2/3} . $$

\noindent
These terms should have following powers

$$ \alpha^2 = \frac{\mu^2}{64 m^4} . $$

\noindent
Therefore the metric does not have physical meaning.

\end{document}